\documentclass[preprint,3p,times]{elsarticle}

\usepackage{lineno,hyperref}
\usepackage{amssymb}
\usepackage{amsmath}
\journal{Physica A}
\date{\today}

\makeatletter
\def\ps@pprintTitle{%
  \let\@oddhead\@empty
  \let\@evenhead\@empty
  \def\@oddfoot{\reset@font\hfill {\footnotesize \@date}}%
  \let\@evenfoot\@oddfoot}
\makeatother

\def\Eq#1{Eq.~(\ref{#1})}
\def\Eqs#1{Eqs.~(\ref{#1})}
\def\Fig#1{Fig.~\ref{#1}}
\def\nn{\nonumber\\}

\def\>{\rangle}
\def\<{\langle}
\def\half{\tfrac{1}{2}}
\def\d{\partial}

\def\Iii{\int_{-\infty}^{\infty}}
\def\tr{\text{tr}}
\def\adg{a^\dagger}
\def\dg{\dagger}
\def\al{\alpha}
\def\bt{\beta}
\def\lam{\lambda}
\def\kap{\kappa}
\def\w{\omega}
\def\del{\delta}
\def\sig{\sigma}
\def\gam{\gamma}
\def\th{\theta}

\def\Del{\Delta}
\def\Gam{\Gamma}
\def\Om{\Omega}

\def\zb{\bar{z}}

\def\DB#1{\boldsymbol{D}(#1)}

\def\LqB{\boldsymbol{L}_0}
\def\vB{\boldsymbol{v}}
\def\uB{\boldsymbol{u}}
\def\VB{\boldsymbol{V}}
\def\UB{\boldsymbol{U}}

\def\Gv{\vec{\Gamma}}
\def\gv{\vec{g}}
\def\thv{\vec{\theta}}
\def\etv{\vec{\eta}}
\def\ev{\vec{e}}

\def\st{\text{st}}
\def\L{\mathcal{L}}
\def\T{\text{T}}
\def\aq{\hat{a}_\text{q}}
\def\acl{\hat{a}_\text{cl}}

\begin{document}

\begin{frontmatter}

\title{Evolution of Gaussian mixed states under the Markovian master equation for a driven quantum oscillator}

\author[1]{B. A. Tay}
\ead{BuangAnn.Tay@nottingham.edu.my}
\address[1]{School of Computer and Mathematical Sciences, Faculty of Science and Engineering, University of Nottingham Malaysia, Jalan Broga, 43500 Semenyih, Selangor, Malaysia}

\begin{abstract}
We study a generic quantum Markovian master equation for a linearly displaced or driven harmonic oscillator. 
It was known that the displacement dynamics of Gaussian mixed states depends on the unitary part of the Liouvillian, the decay rate of the system but not on the bath temperature.
Here we further show that the fast-rotating modes do not affect the system's displacement dynamics under linear driving forces. 
Analytical solutions of the quantum master equation are obtained for displaced Gaussian mixed states. 
Because the non-driven and driven Liouvillians are related by a unitary displacement operator, they are expected to share the same exceptional points structure. 
At the exceptional points, the displacement of critically damped oscillator displays a characteristics polynomial-in-time prefactor multiplied by an exponential decay.
We discuss how external time-dependent forces affect the displacement dynamics using impulsive force and harmonic force as examples.
The results obtained for constant driving remain valid in the presence of time-dependent driving.
\end{abstract}


\end{frontmatter}

\section{Introduction}

Gaussian states play a central role in quantum optics \cite{Walls08}, continuous variable quantum information \cite{Nielsen,Adesso14} and in the studies of quantum decoherence \cite{Decoh03}. They can be generated in the laboratory, in general remain Gaussian under dissipation and decoherence, and are fully characterized by the first and second moments \cite{Simon87,GardinerStoch,Serafini23}. Depending on the context, they may be represented in various equivalent forms, such as the Wigner functions \cite{Hillery84,Balazs84}, the Glauber-Sudarshan P representation \cite{Klauder68,GardinerQNoise}, the Q representation \cite{GardinerQNoise}, or representations in terms of Gaussian operators \cite{Englert03}. 
Common examples include vacuum, coherent, squeezed, and thermal states.  

Displacement of Gaussian states can be implemented experimentally using external driving fields \cite{Carruthers65,Klauder68,GardinerQNoise}. Coherent states, squeezed coherent states, and thermal coherent states are examples of displaced Gaussian states. In the laboratories, they can be produced across multiple platforms, including laser-driven optical cavities \cite{Barlow15}, optomechanical resonators \cite{Wilson-Rae08}, externally driven microwave cavities \cite{Portugal23}, laser-driven trapped atoms \cite{McCormick19}, and etc.

In view of the importance of Gaussian states, their mathematical properties \cite{Simon87,GardinerStoch,Serafini23} and their evolution under the influence of environment have been extensively investigated. Classically, their evolution is governed by the Fokker-Planck equation \cite{Risken}. In quantum systems, the Gorini-Kossakowski-Sudarshan-Lindblad (GKSL) Markovian master equations \cite{Kossa76,Lindblad76} are widely used to model dissipation and decoherence of Gaussian states induced by environment \cite{GardinerQNoise,Breuer,Decoh03,Lu03,Yang03}. 
Most prior studies focused either on non-driven Gaussian states \cite{Breuer,Decoh03}, isolated but driven oscillators \cite{GardinerQNoise}, or with initial Gaussian pure states \cite{Lu03,Yang03}.

In this work, we consider a generic quantum Markovian master equation (MME) for a linearly driven quantum harmonic oscillator and study its action on displaced Gaussian mixed states.
Based on our earlier works on generic quantum MMEs \cite{Tay17,Tay17b,Tay20,Tay23}, we extend the studies to include linear driving and obtain closed-form solutions for general Gaussian mixed states.
The results are consistent with known results from previous studies \cite{Serafini23,McDonald23} that the displacement dynamics of the driven oscillator depends only on the unitary (Hamiltonian) part of the quantum Liouvillian together with the decay rate carried by the dissipative operator. 
Other environmental properties, such as the bath temperature do not influence the displacement dynamics.
Here we further show that the fast-rotating modes also do no affect the displacement dynamics. 
The results remain valid under external time-dependent driving forces.

One important development in non-Hermitian physics is the recent progress in the studies of non-Hermitian degeneracies or exceptional points when two or more eigenvalues coalesce \cite{Berry04,Heiss12,Ashida21}. The exceptional points in quantum Liouvillian were discussed in Ref.~\cite{Nori19} and their topological aspects in Ref.~\cite{Bergholtz21}. 
For a quantum oscillator, its exceptional points structure was obtained for a generic Liouvillian \cite{Tay23}, the multimode extension to GKSL equation \cite{Gaidash25}, and parametrically driven oscillator \cite{Downing23,Mylnikov25}. 
Here we discuss the exceptional points of a driven oscillator for a generic Liouvillian. 
It is known that driven Liouvillian differs from non-driven one by a linear term and is related to it by a displacement transformation \cite{Klauder68,Serafini23}. 
This implies that they share the same exceptional points structure. 
The conclusion is valid for a generic Liouvillian with the fast-rotating modes as well as when it is subjected to time-dependent external driving.
This work should be useful to researchers working in quantum continuous variable systems, for example, quantum information \cite{Weedbrook12}, quantum resonators \cite{Barlow15,Wilson-Rae08,Portugal23,McCormick19}, and synchronization of quantum oscillators under external time-dependent driving \cite{Walter14,Li25}.

Other closely related developments in continuous-variable open quantum systems include the use of symmetries of the reduced dynamics
to determine the spectrum of Liouvillians \cite{Tay20,McDonald23,Gaidash25} and the exceptional point structures \cite{Tay23,Gaidash25} in single-  and multi-mode systems. 
Through thermal or dissipative symmetries, the bath temperature and diffusion coefficients in the reduced dynamics of single- \cite{Tay04,Tay07,Tay17} and multi-mode systems \cite{McDonald23,Gaidash25} can be transformed or gauged accordingly. This provides a plausible explanation for the independence of the first moments from on the bath temperature or noise \cite{Serafini23,McDonald23}, even though systems related by thermal symmetries are physically distinct in principle. 
Here we extend the explanation to encompass the fast-rotating modes of reduced dynamics. 
Recently, linear driving for continuous variable open quantum systems were explored in Refs.~\cite{Vazquez18,Ma18}.

The organization of this paper is as follows.
In Section \ref{SecNonDisplace}, we review earlier results on generic quantum MME and express the Liouvillian in terms of the familiar GKSL operator.
Section \ref{SecDisplace} introduces the displacement superoperator and describes its action on Gaussian mixed states.
In Section \ref{SecSolMME}, we consider the generic Liouvillian of quantum MME for a driven oscillator, and obtain its solutions for Gaussian mixed states.
Section \ref{SecEP} discusses the Liouvillian exceptional points of the system.
The effects of time-dependent driving on the displacement dynamics are discussed using the impulsive and harmonic external forces in Section \ref{SecTmDep}. 
We summarize the main results in Section \ref{SecConclusion}.
Details on the position representation of superoperators, solutions to a non-Hermitian eigenvalue problem, solutions to the MME and etc., are given in the appendices.

\section{Markovian master equation for non-driven oscillator}
\label{SecNonDisplace}

We consider the quantum MME \cite{Tay17,Tay17b}
\begin{align}   \label{drho}
   \frac{\d\rho}{\d t}=\L_0\rho\,,
\end{align}
with a generic quantum Liouvillian,
\begin{align}   \label{Liouv0}
    \L_0\rho&=-i[H_0,\rho]-\frac{1}{2}(\eta_0-\gam)L\rho
            -\frac{1}{2}(\eta_0+\gam)R\rho
            -\chi V\rho-\chi^*V^\dg\rho\,.
\end{align}
The superoperators are quadratic in the creation $\adg$ and annihilation $a$ operators of the oscillator.
The Hamiltonian is
\begin{align}   \label{H0}
    H_0&=\w_0 \adg a+i\frac{1}{2}(\xi^* a^2-\xi{\adg}^2)\,,
\end{align}
where $\w_0$ is the natural frequency of the oscillator.
We use the units $\hbar=1$.
The second term in \Eq{H0} is the generator of the squeezed operator \cite{GardinerQNoise}, with a complex parameter
\begin{align}   \label{xi}
    \xi&\equiv \frac{1}{2}(\th_2+i\th_1)\,.
\end{align}
In \Eq{Liouv0}, $\chi$ is the complex parameter
\begin{align}   \label{chi}
    \chi&\equiv \frac{1}{2}(\eta_1+i\eta_2)\,.
\end{align}

We note that in our previous works \cite{Tay07,Tay08,Tay17,Tay17b,Tay19b,Tay20,Tay23} where we explored the transformation properties and solutions of quantum Liouvillian, it was more convenient to use a different set of superoperators,
\begin{align}   \label{Kgen}
    K_0
    &=-\L_0=2\w_0 iL_0+\th_1iM_1+\th_2iM_2+\gam(O_0-I/2)+\eta_0O_++\eta_1L_{1+}+\eta_2L_{2+}\,,
\end{align}
from which the definitions of the parameters in $\L_0$ originate from, cf.~\Eqs{xi} and \eqref{chi}.
The superoperators are listed in \ref{AppRepres} both in the creation-annihilation operators and coordinate representation for the convenience of the readers.

How the parameters affect the displacement or first moments as well as the exceptional points will be discussed later in Section \ref{SecLdisp} and Section \ref{SecEP}, respectively. We first discuss their effects on the second moments.
In terms of this set of superoperators, the actions of $e^{-i\th_1M_1}, e^{-i\th_2M_2}$ on quadratic operators produce squeezing in the second moments of the phase space variables \cite{GardinerQNoise} defined by
\begin{align} \label{2ndmoment}
\sig_{\!\!xx}=\<\hat{x}^2\>-\<\hat{x}\>^2\,,\qquad \sig_{\!\!pp}=\<\hat{p}^2\>-\<\hat{p}\>^2\,,\qquad \sig_{\!\!xp}=\frac{1}{2}\<\hat{x}\hat{p}+\hat{p}\hat{x}\>-\<\hat{x}\>\<\hat{p}\>\,,
\end{align}
where $\<\hat{o}\>\equiv\tr(\hat{o}\rho)$. 
The position and momentum operators are
\begin{align}   \label{xhat}
\hat{x}=\frac{1}{\sqrt{2}}(\adg+a)\,, \qquad \hat{p}=\frac{i}{\sqrt{2}}(\adg-a)\,,
\end{align}
where we have used dimensionless position and momentum.
The corresponding coordinates with dimensions of length $x_d$ and momentum $p_d$ are related to the dimensionless ones by
\begin{align}   \label{xpDless}
x=x_d\sqrt{\frac{m\w_0}{\hbar}}\,, \qquad p=\frac{p_d}{\sqrt{m\hbar\w_0}}\,,
\end{align}
respectively.
Defining the squeezed moments by a prime, such as $\<\hat{o}'\>=\<e^{-i\th_1M_1}\hat{o}\>$, the action of $e^{-i\th_1M_1}$ produces the following squeezing in the variances \cite{Tay19b}
\begin{align} 
    \left(\begin{array}{c}
            \frac{1}{2}(\sig'_{\!\!xx}+\sig'_{\!\!pp}) \\
            \sig'_{\!\!xp}
          \end{array}\right) =\left(\begin{array}{cc}
           \cosh\th_1 & \sinh\th_1 \\
           \sinh\th_1 & \cosh\th_1 
          \end{array}\right)\left(\begin{array}{cc}
           \frac{1}{2}(\sig_{\!\!xx}+\sig_{\!\!pp}) \\
            \sig_{\!\!xp}
          \end{array}\right)\,,
\end{align}
while $\sig'_{\!\!xx}-\sig'_{\!\!pp}=\sig_{\!\!xx}-\sig_{\!\!pp}$ remains invariant, whereas $e^{-i\th_2M_2}$ produces the squeezing $\sig'_{\!\!xx}=e^{\th_2}\sig_{\!\!xx}$, $\sig'_{\!\!pp}=e^{-\th_2}\sig_{\!\!pp}$, whereas $\sig'_{\!\!xp}=\sig_{\!\!xp}$ is unaffected.

Furthermore, some suitable unitary transformations \cite{Tay20} with the generators $iL_0, iM_1, iM_2$ can always diagonalize $H_0$ into the form $H'_0=\w\adg a$, where
\begin{align}   \label{w}
    \w&\equiv\sqrt{\w_0^2-\frac{\th_1^2}{4}-\frac{\th_2^2}{4}}\,.
\end{align}
We shall call $\w$ the renormalized frequency of the oscillator.

In \Eq{Liouv0}, $L$ and $R$ are the dissipative operators \cite{Kossa76,Lindblad76},
\begin{align}   \label{Lrho}
    L\rho&=a\rho\adg-\frac{1}{2}\adg a\rho -\frac{1}{2}\rho \adg a\,,\qquad
    R\rho=\adg\rho a-\frac{1}{2}a\adg \rho -\frac{1}{2}\rho a \adg\,.
\end{align}
$L$ describes the relaxation of higher excited states to lower ones, and $R$ describes the reverse process of excitation from lower states to higher ones.
$\gam$ is the decay rate of the system. 
$\eta_0$ is a parameter related to the temperature of the thermal bath. In the GKSL equation \cite{Kossa76,Lindblad76} and the Caldeira-Leggett (CL) equation \cite{CL83}, $\eta_0=-\gam(2\bar{n}+1)$, where $\bar{n}=1/(\exp(\w_0/k_BT)-1)$ is the average occupation number of the bath modes, $k_B$ is the Boltzmann constant and $T$ is the temperature of a bath.

To understand how $L$ and $R$ affect the second moments, it is more convenient to use the equivalent set of superoperators $(O-I/2)$ and $O_+$, see \Eq{O0}. The effects of $e^{\al(O-I/2)}$ and $e^{\eta_0 O_+}$ are \cite{Tay19b}
\begin{subequations}
\begin{align} \label{O02nd}
   \sig'_{\!\!xx}&=e^\al \sig_{\!\!xx}\,, \qquad\quad \sig'_{\!\!pp}=e^\al \sig_{\!\!pp}\,, \qquad\quad \sig'_{\!\!xp}=e^\al \sig_{\!\!xp}\,,\\
   \sig'_{\!\!xx}&=\sig_{\!\!xx}+\frac{\eta_0}{2} \,, \qquad \sig'_{\!\!pp}=\sig_{\!\!pp}+\frac{\eta_0}{2} \,, \qquad \sig'_{\!\!xp}=\sig_{\!\!xp}\,.\label{O+2nd}
\end{align}
\end{subequations}
These relations suggest that the effects of the diffusion coefficients could be transformed or gauged away by $e^{\al(O-I/2)}$ and $e^{\eta_0 O_+}$.
This forms the basis of thermal symmetry \cite{Tay07} and dissipative symmetry \cite{McDonald23} that make use of $e^{\al(O-I/2)}$ and $e^{\eta_0 O_+}$, respectively.
In similar way, the symmetry had been generalized to include $L_{1+}$ and $L_{2+}$ as well \cite{Tay17}.

The $V$ and $V^\dg$ involves the so-called `virtual' transitions \cite{Passante95},
\begin{align}   \label{Vrho}
    V\rho&=a\rho a-\frac{1}{2}a a\rho -\frac{1}{2}\rho aa\,,\qquad
    V^\dg\rho=\adg\rho \adg-\frac{1}{2}\adg\adg\rho -\frac{1}{2}\rho \adg\adg\,, 
\end{align}
with a strength determined by $\chi$ \eqref{chi}.
They come from the fast-rotating bath modes that are usually dropped from the Liouvillian $\L_0$ by imposing the rotating-wave approximation \cite{GardinerQNoise,Breuer,Walls08}.
The $V, V^\dg,$ or their counterparts $L_{1+}, L_{2+}$ \eqref{L2+a}, are the supergenerators of the fast-rotating modes.
The $\eta_1$-term appears in the CL equation, where $\eta_1=\eta_0$, which belongs to the diffusive modes of the reduced dynamics, cf.~the coordinate representations of $O_+$ and $L_{1+}$ in \Eq{L2+}.
On the other hand, $\eta_2$-term appears in the Hu-Paz-Zhang (HPZ) equation \cite{HPZ92} under general environment.
This so-called anomalous diffusion term will not lead to the usual Pauli-type kinetic equation \cite{Barsegov02}.
The effects of $e^{\eta_1L_{1+}}$ and $e^{\eta_2L_{2+}}$ on the second moments are \cite{Tay19b}
\begin{subequations}
\begin{align}   \label{L1+2nd}
   \sig'_{\!\!xx}&=\sig_{\!\!xx}-\frac{\eta_1}{2} \,, \qquad \sig'_{\!\!pp}=\sig_{\!\!pp}+\frac{\eta_1}{2} \,, \qquad \sig'_{\!\!xp}=\sig_{\!\!xp}\,,\\
   \sig'_{\!\!xx}&=\sig_{\!\!xx} \,, \qquad\qquad \sig'_{\!\!pp}=\sig_{\!\!pp} \,, \qquad\qquad \sig'_{\!\!xp}=\sig_{\!\!xp}+\frac{\eta_2}{2} \,.   \label{L2+2nd}
\end{align}
\end{subequations}

We also consider solutions of the MME \eqref{drho}-\eqref{Liouv0} for Gaussian mixed states \cite{Tay17b}. For this purpose, it is more convenient to work in the coordinate representation, in which a density operator is denoted by
\begin{align}   \label{rhoxy}
    \rho(Q,r)&\equiv\left\<Q+\frac{r}{2}\bigg|\rho\bigg|Q-\frac{r}{2}\right\>
    =\<x|\rho|y\>\,,
\end{align}
where we introduce the center and the relative coordinates \cite{Wigner32},
\begin{align} \label{Qrxx}
    Q&\equiv\frac{1}{2}(x+y)\,,   \qquad r\equiv x-y\,.
\end{align}
In these coordinates, generic Gaussian states take the form
\begin{align}   \label{rhog}
    \zeta_0(Q,r,t)&=\sqrt{\frac{2\mu(t)}{\pi}}\exp\left(-2\mu(t) Q^2-i\kap(t) Qr-\half(\mu(t)+\nu(t))r^2\right)\,,
\end{align}
where $\mu(t), \nu(t)$ and $\kap(t)$ are time-dependent real parameters.
By definition, Gaussian pure states $|\psi\>$ are separable in the $x$ and $y$ coordinates, $\<x|\psi\>\<\psi|y\>$.
This implies $\nu=0$.
With the positive-semidefiniteness requirement on Gaussian states \cite{Tay17b}, $\mu>0$ and $\nu\geq0$, $\zeta_0$ are Gaussian mixed states whenever $\nu>0$.
The form of Gaussian states \eqref{rhog} is preserved under the evolution governed by the Liouvillian $\L_0$.
The time evolution of $\mu, \nu, \kappa$ and their properties were investigated in Ref.~\cite{Tay17b}. 
They are listed in \ref{AppSol} together with their stationary state expressions for the convenience of the readers.

It is known that Gaussian states are fully characterized by the first and second moments \cite{Simon87,GardinerStoch,Englert03,Serafini23}. 
We use the subscript `0' on $\zeta_0$ to denote the fact that it has zero average position and momentum, 
\begin{align}
\<x(t)\>_0=\<p(t)\>_0=0\,,
\end{align}
i.e., it is a non-displaced Gaussian state.
The subscript `0' on $\<\cdot\>_0$ denotes the trace of the operators with $\zeta_0$, for example, $\<x\>_0=\tr(\hat{x}\zeta_0)$, and so on.
The second moments for the state are \cite{Tay17b},
\begin{align}   \label{sigxx}
    \sig_{xx,\,0}(t)=\frac{1}{4\mu(t)}\,,
    \qquad
    \sig_{pp,\,0}(t)=\frac{\Del^2(t)}{4\mu(t)}\,,\qquad
    \sig_{xp,\,0}(t)=-\frac{\kap(t)}{4\mu(t)}\,,
\end{align}
cf.~the definition in \Eq{2ndmoment}.

The Wigner representation with phase-space-like coordinates are useful in many applications \cite{GardinerQNoise,Walls08,Serafini23}.
The Wigner function is obtained through a Fourier transform on the $r$ coordinate \cite{Hillery84,Balazs84},
\begin{align}   \label{W}
   W(Q,P)\equiv\frac{N}{2\pi}\Iii dr e^{-iPr}\rho(Q,r)\,,
\end{align}
where $N$ is a normalization constant that ensures
\begin{align}   \label{Wnorm}
    \Iii dQ\Iii dP \, W(Q,P)=1\,.
\end{align}
In the Wigner representation, the Gaussian state $\zeta_0$ has the form
\begin{align}   \label{Wignerrho}
    W_0(Q,P,t)&=\frac
    {1}{\pi}\sqrt{\frac{\mu(t)}{\mu(t)+\nu(t)}} \exp\left(-\frac{\Del^2(t) Q^2+2\kap(t) QP+P^2}{2(\mu(t)+\nu(t))}\right)\,,\\
    \Del^2(t)&=4\mu(t)\big(\mu(t)+\nu(t)\big)+\kap^2(t)\,.
\end{align}

\section{Displaced Gaussian mixed states}
\label{SecDisplace}

In quantum mechanics, a driven oscillator is described by the coherent state $|z\>=u(z)|0\>$ \cite{Klauder68,GardinerQNoise}, which can be generated by acting the displacement operator
\begin{align}   \label{u}
   u(z)\equiv e^{z\adg-z^*a}
\end{align}
on the vacuum state $|0\>$. 
Below, we  recall some known properties of the displacement operator \cite{Klauder68,GardinerQNoise,Walls08,Serafini23} and introduce notations.
The coherent state has a displaced position $\<z|\hat{x}|z\>=q$ and momentum $\<z|\hat{p}|z\>=p$ encoded in the complex parameter
\begin{align}   \label{z}
    z&\equiv\frac{1}{\sqrt{2}}(q+ip)\,.
\end{align}
The corresponding displacement superoperator in the quantum Liouville space is
\begin{align}   \label{eD}
    e^{D(z)}\rho&=u(z)\rho u^\dg(z)\,,
\end{align}
where $D(z)$ is the generator of the displacement superoperator 
\begin{align}   \label{D}
    D(z)\rho&\equiv (z\adg-z^* a)\rho-\rho (z\adg-z^* a)\,.
\end{align}

In \ref{AppDform} we show that (i) $D(z)$ is adjoint-symmetric $\tilde{D}(z)=D(z)$ \cite{Prigogine73,Tay17}, which implies that $\exp(D(z))$ preserves the hermiticity of the density matrix. 
$D(z)$ is also (ii) trace-preserving, $\tr(D(z)\rho)=0$, so that it conserves probability.
The proof in the appendix makes use of cyclic permutations of operators under a trace, which requires $\rho$ to be differentiable and vanishes fast enough at infinity, see \ref{AppMixSt} for the details.
These properties are satisfied by the Gaussian states.
It is also shown in \ref{AppDform} that \Eq{D} is the generic form of operators linear in the position coordinates and simultaneously satisfy the two conditions \eqref{Gs}.
In the position coordinates, $D(z)$ becomes \eqref{DQr}
\begin{align}   \label{Dx}
   D(z)=ipr-q\frac{\d}{\d Q}\,.
\end{align}

To obtain the Gaussian mixed states of a displaced oscillator, denoted by $\zeta_z$, we act the displacement superoperator \eqref{eD} on $\zeta_0$ \eqref{rhog}. 
Exponentiation of the differential operator $-q\d/\d Q$ displaces the coordinate $Q$ of the Gaussian mixed states to $Q-q$. 
On the other hand, exponentiation of $ipr$ multiplies an overall phase to the Gaussian mixed states.
As a result, we obtain
\begin{align}   \label{rhod}
    \zeta_z(Q,r,t)\equiv e^{D(z)}\zeta_0(Q,r,t)
    =e^{ipr}\zeta_0(Q-q,r,t)\,.
\end{align}
Therefore, the application of the displacement operator on the Gaussian states changes the displacement in the $Q$ coordinates and add a phase factor in the $r$ coordinate.
It does not affect the evolution of the parameters $\mu, \kappa$ and $\nu$ in the Gaussian states, which determine the second moments \eqref{sigxx}.
This is the standard separation between the drift and diffusion processes in stochastic processes \cite{Breuer,GardinerStoch,Serafini23}.
The effects of the displacement superoperator on the Gaussian mixed states are completely known once we determine $q$ and $p$.  

We calculate the average position and average momentum of $\zeta_z$ as follows,
\begin{align}   \label{avexrho}
   \<x(t)\>_z&=\tr[\hat{x}\zeta_z(t)] \nn
   &=\tr\big[\hat{x}e^{D(z)}\zeta_0(t)\big] \nn
   &=\tr\big[\hat{x}u(z)\zeta_0(t)u^\dg(z)\big] \nn
   &=\tr\big[u^\dg(z)\hat{x}u(z)\zeta_0(t)\big]\,,
\end{align}
after using the definition of the displacement superoperator \eqref{eD} and the cyclic permutation of operators under the trace.
Then, we expand the expression
\begin{align}   \label{aau}
   u^\dg(z)\hat{x}u(z)&=e^{-z\adg+z^* a}\hat{x}e^{z\adg-z^* a} =\hat{x}-[z\adg-z^* a,\hat{x}]+\frac{1}{2!}[z\adg-z^* a,[z\adg-z^* a,\hat{x}]]+\cdots\,,
\end{align}
and use $\hat{x}$ from \Eq{xhat} to calculate the commutator $[z\adg-z^* a,a+\adg]=-z-z^*=-\sqrt{2}q$. 
Consequently, $[z\adg-z^* a,[z\adg-z^* a,\hat{x}]]=0$, i.e., the rest of the commutators in \Eq{aau} vanishes. 
We conclude that the average position is displaced by $q$,
\begin{align}   \label{avexrhofinal}
   \<x(t)\>_z&=\<x(t)+q\>_0=\<x(t)\>_0+q\,.
\end{align}
Using $\hat{p}$ from \Eq{xhat}, we can similarly show that the average momentum is displaced by $p$,
\begin{align}   \label{aveprho}
   \<p(t)\>_z&=\<p(t)+p\>_0=\<p(t)\>_0+p\,.
\end{align}
This is also evident once we calculate the Wigner function of $\zeta_z$ using \Eq{W},
\begin{align}   \label{Wignerrhod}
    W_z(Q,P,t)&=\frac
    {1}{\pi}\sqrt{\frac{\mu(t)}{\mu(t)+\nu(t)}} \exp\left(-
    \frac{\Del(t)^2 (Q-q)^2+2\kap(t) (Q-q)(P-p)+(P-p)^2}{2(\mu(t)+\nu(t))}
    \right)\,,
\end{align}
which is explicitly displaced in both of the phase space coordinates.
The second moments of the displaced Gaussian state remains the same as in \Eq{sigxx}. This can be shown, for instance,
\begin{align}   \label{avexxz}
   \<x^2(t)\>_z&=\tr[\hat{x}^2\zeta_z(t)] \nn
   &=\tr\big[(u^\dg(z)\hat{x}u(z))^2\zeta_0(t)\big] \nn
   &=\tr\big[(\hat{x}+q)^2\zeta_0(t)\big] \nn
   &=\<x^2(t)\>_0+2q\<x(t)\>_0+q^2\,.
\end{align}
Then,
\begin{align}   \label{sigxxz}
   \sig_{xx,\,z}&=\<x^2(t)\>_z-\<x(t)\>^2_z \nn
   &=\<x^2(t)\>_0+2q\<x(t)\>_0+q^2-(\<x(t)\>_0+q)^2\nn
   &=\sig_{xx,\,0}
\end{align}
We can likewise show that $\sig_{pp,\,z}=\sig_{pp,\,0}$ and $\sig_{xp,\,z}=\sig_{xp,\,0}$.

In summary, $\zeta_z$ is displaced Gaussian state with its center driven to $q$ and $u$ in the position and momentum space, respectively. It is the mixed state generalization of the pure coherent state $|z\>\<z|$.

\section{Solutions to quantum Markovian master equation for driven oscillator}
\label{SecSolMME}

\subsection{Liouvillian for driven oscillator}
\label{SecLdisp}

In this section we first obtain the generic quantum MME for driven oscillator \cite{Klauder68,GardinerQNoise} using the displacement superoperator. Then we consider its solutions for Gaussian mixed states.

Applying the displacement superoperator $\exp(D(z))$ to equation \eqref{drho}, we obtain the MME
\begin{align}   \label{drhoD}
    \frac{\d \tau}{\d t}=\L_z\tau\,,
\end{align}
where to avoid confusion with the MME for non-driven oscillator \eqref{drho},  we have labeled the density operator by $\tau$.
In the coordinate representation, it is
\begin{align}   \label{tau}
    \tau(Q,r,t)&\equiv e^{D(z)}\rho(Q,r,t)\,,
\end{align}
where $\rho(t)=\exp(\L_0 t)\rho(0)$ is the solution to the MME for non-driven oscillator \eqref{drho}.
The generator of time evolution is
\begin{align}   \label{K'eq}
    \L_z&\equiv e^{D(z)}\L_0e^{-D(z)} 
    =\L_0+[D(z),\L_0]+\frac{1}{2!}[D(z),[D(z),\L_0]]+...\,.
\end{align}
Using the commutation relations listed in \Eqs{L0comm}-\eqref{O0comm}, we find that
\begin{align}   \label{commDK}
    [\L_0,D(z)]=-D(\al)\,,\qquad
    [D(z),D(\al)]=0\,,
\end{align}
with the parameter
\begin{align}   \label{alz}
    \al\equiv\frac{1}{\sqrt{2}}(\al_q+i\al_p)\,,
\end{align}
where
\begin{subequations}
\begin{align}   \label{alq}
    \al_q&=\half(\gam+\th_2)q-\half(2\w_0-\th_1)p\,,\\
    \al_p&=\half(2\w_0+\th_1)q+\half(\gam-\th_2)p\,.\label{alp}
\end{align}
\end{subequations}
As a result, the Liouvillian acquires a displaced term,
\begin{align}   \label{Kz}
    \L_z&=\L_0+D(\al)\,.
\end{align}
Notice that $\al$ is the coefficient of the displacement superoperator in the MME, whereas $z$ is the actual displacement induced on the oscillator's Gaussian state.
We can determine the actual displacement of the Gaussian state, $z$ \eqref{z}, by inverting \Eqs{alq}-\eqref{alp},
\begin{subequations}
\begin{align}   \label{qalq}
    q&=\frac{2}{\gam^2+4\w^2}\left((\gam-\th_2)\al_q+(2\w_0-\th_1)\al_p\right)\,, \\
    p&=\frac{2}{\gam^2+4\w^2}\left(-(2\w_0+\th_1)\al_q+(\gam+\th_2)\al_p\right)\,,
    \label{palp}
\end{align}
\end{subequations}
where $\w$ is the renormalized frequency of the oscillator \eqref{w}.
From the expressions in \Eqs{qalq}-\eqref{palp}, we learn that the actual displacement of the Gaussian states is affected by the coefficients of the part of Liouvillian that do not commute with the displacement superoperator, i.e.~the unitary part $H_0$ and the dissipative part $R-L$ with coefficient $\gam$ in $\L_0$ \eqref{Liouv0}, cf.~\Eqs{L0comm}-\eqref{O0comm}, or equivalently, the coefficients of $iL_0, iM_1, iM_2$ and $(O_0-I/2)$ in $K_0$ \eqref{Kgen}, cf.~\Eqs{iL0a}-\eqref{L2+a}.
On the contrary, the influence of $R+L$, $V$ and $V^\dg$, or equivalently, $O_+, L_{1+}$ and $L_{2+}$, that commute with the displacement superoperator are absent from the actual displacement.
The results are consistent with the results in Ref.~\cite{Ma18}, indicating a separation between the drift and diffusion processes \cite{GardinerStoch,Breuer,Serafini23}.

The generality of this conclusion can also be seen by evaluating the time derivative of the first moment. This requires the transposition of superoperators defined in \ref{App1stmoment}, where we show that the transposition of the superoperators $O_+, L_{1+}, L_{2+}$ annihilate linear operators $\hat{x}, \hat{p}$, for instance, $O_+^\T(\hat{x})=0$ and etc. 
We refer the readers to \ref{App1stmoment} for further details. As a result, $\eta_0=-\gam(2\bar{n}+1)$ that depends on the bath temperature and the coefficients of the fast-rotating modes $\eta_1, \eta_2$ drop out from the equation of motion of the first moments. 
We know that $\L_0$ is the most general form of Markovian Liouvillian in the Liouville space of an oscillator \cite{Talkner81,Tay17}. In the non-Markovian case, the most general Liouvillian has the same generic form except that the coefficients may be time-dependent, see the Hu-Paz-Zhang equation \cite{HPZ92} for example. Therefore, the conclusion that the bath temperature and fast-rotating modes do not contribute to the dynamics of the first moment applies to the non-Markovian case as well.
Since a time-dependent external force can be represented by linear operators, as discussed in Section \ref{SecTmDep}, the results remain valid even when such a force is applied to the oscillator.

Other way to explain the independence of first moments from the diffusive coefficients is through the symmetries of reduced dynamics \cite{Tay07,Tay17,McDonald23}, where the diffusive coefficients $\eta_0, \eta_1, \eta_2$ can be transformed or gauged through unitary or similarity transformations.
In the thermal symmetry of reduced dynamics, a unitary but non-trace preserving superoperator $\exp(iG\th)$, where $G=2iO_0$ and real $\th$, can transform the bath temperature in the GKSL equation or in the CL equation to zero \cite{Tay07}.
Thermal symmetry has been generalized using the symplectic algebra \cite{Gilmore,Simon87,Simon88} to include $O_+, L_{1+}$ and $L_{2+}$. The result is a set of trace-preserving superoperators $e^{\al (O_0-I/2)}, e^{\al O_+}, e^{\al L_{1+}}$ and $e^{\al L_{2+}}$ with real $\al$ that act on the Liouvillian through similarity transformation \cite{Tay17}.
In the dissipative transformation \cite{McDonald23}, the supergenerator $O_+$ (in our notations, see \ref{Appa} for the details) is used to gauge away noise and the results extended to multimode systems.
Though the symmetry arguments provide a plausible explanation for the observed independence, it should be emphasized that systems connected by similarity transformation are physically distinct, even though they share certain dynamical properties.

We note that thermal symmetry has been extended to two-level and $N$-level systems \cite{Tay19a}. In the two-level case, the diffusive parts of the amplitude damping Liouvillian \cite{Nielsen} do not annihilate the spin operators or the Pauli matrices $\sig_i$. Hence, the time evolution of the Bloch vector $(\<\sig_x\>\,\, \<\sig_y\>\,\, \<\sig_z\>)^\T$ depends on the bath temperature.
Since finite-level systems are outside the scope of this work, we will not discuss it further.

\subsection{Solutions to MME for non-driven oscillator}
\label{SecInidisp}

To obtain the solution to $\tau(t)$, it is a good preparation to first consider the solution to the MME for non-driven oscillator \eqref{drho} with initially displaced Gaussian mixed states.
The solutions we obtained in Ref.~\cite{Tay17b} for the MME \eqref{drho} uses the initial condition of non-displaced Gaussian mixed states $\zeta_0(Q,r,0)$. When the initial condition is a Gaussian mixed state displaced by $z_0$, the initial state is
\begin{align}   \label{inirhoz0}
    \rho(Q,r,0)=e^{D(z_0)}\zeta_0(Q,r,0)=\zeta_{z_0}(Q,r,0)\,,
\end{align}
where we use \Eq{rhod} for the definition of the subscript on $\zeta_{z_0}$.
The solution to the MME \eqref{drho} is (omitting the coordinates dependence for simplicity)
\begin{align}   \label{rhodt}
    \rho_{\zb}(t)&\equiv e^{\L_0t}\rho(0)\nn
    &=e^{\L_0t}e^{D(z_0)}e^{-\L_0t}\cdot e^{\L_0t}\zeta_0(0)\nn
    &\equiv e^{D(\zb_0(t))}\zeta_0(t)\nn
    &=\zeta_{\zb_0(t)}(t)\,,
\end{align}
where the displacement parameter $\zb(t)$ is defined through the relation
\begin{align}   \label{KDK}
    D(\zb_0(t))&\equiv e^{\L_0t}D(z_0)e^{-\L_0t}\,.
\end{align}
To obtain $\zb_0(t)$, we need to solve a non-hermitian eigenvalue problem. The details are worked out in \ref{AppEigevProb}. 
Writing $\zb_0\equiv(\bar{q}_0+i\bar{p}_0)/\sqrt{2}$, the results are
\begin{subequations}
\begin{align}   \label{qb}
    \bar{q}_0(t)&=e^{-\gam t/2}\left(q_0\cos(\w t)
        -\half\big[\th_2 q_0-(2\w_0 -\th_1)p_0\big]\frac{\sin(\w t)}{\w}\right)\,,\\
    \bar{p}_0(t)&=e^{-\gam t/2}\left(p_0\cos(\w t)
        -\half\big[(2\w_0 +\th_1)q_0-\th_2 p_0\big]\frac{\sin(\w t)}{\w}\right)\,.   \label{pb}
\end{align}
\end{subequations}
$\bar{q}_0(t)$ and $\bar{p}_0(t)$ are the average displaced position and momentum of the corresponding Gaussian mixed state caused by the initial displacement.
The state executes oscillatory motion with diminishing amplitude, which eventually settles down to a stationary state with zero average position and momentum, i.e., $\rho_{\zb}(Q,r,t\rightarrow\infty)=\zeta_0(Q,r,0)$.

\subsection{Solutions to MME for driven oscillator}
\label{SecKalInidisp}

With an initially displaced Gaussian mixed state $\tau(Q,r,0)=e^{D(z_0)}\zeta_0(Q,r,0)$, the solution to the MME for driven oscillator \eqref{drhoD} is $\tau(t)=\exp(\L_z t)\tau(0)$. We could have tried to break $\exp(\L_zt)$ into a product of exponentials using the inverse of the Bakers-Campbell-Hausdorff (BCH) formula, i.e., the Zassenhaus formula \cite{Magnus54}. 
The decomposition will have the form,
\begin{align} \label{Zassenhaus}
    e^{\L_zt}&=e^{\L_0 t} e^{D(\al t)}
    e^{D(\al' t^2/2)} e^{D(\al'' t^3/6)}\cdots
        =e^{\L_0 t}e^{D(\al t+\al' t^2/2+\al'' t^3/6+\cdots)}\,,
\end{align}
where to obtain the last equality, we sum the arguments of $D$ based on the fact that the $D$s with different arguments commute \eqref{commDK}. However, this approach is not useful because the sum cannot be put in a simple form.

We instead decompose the time evolution for an initially displaced $\tau(0)$ as follows,
\begin{align}   \label{taut}
    \tau(t)&=e^{\L_z t}\tau(0)\nn
        &=e^{D(z)}e^{\L_0 t}e^{-D(z)}\cdot e^{D(z_0)}\zeta_0(0)\nn
        &=e^{D(z)}\cdot e^{\L_0 t}e^{D(z_0-z)}e^{-\L_0 t}\cdot e^{\L_0 t}\zeta_0(0)\nn
        &=e^{D(z)}e^{D(\overline{z_0-z}(t))}\zeta_0(t)\nn
        &=e^{D(z+\overline{z_0-z}(t))}\zeta_0(t)\nn
        &\equiv\zeta_{\bt(t)}(t)\,,
\end{align}
where in the last line we have defined the displacement
\begin{align}   \label{bt}
    \bt(t)\equiv z+\overline{z_0-z}(t)\,,
\end{align}
and use the definition \eqref{inirhoz0}, where $\zeta_0(t)$ in \Eq{taut} is the already known solution to the MME for non-driven oscillator \eqref{drho}. 
Therefore, we can make use of the property of displacement superoperator to obtain the total displacement.
Using \Eqs{qb} and \eqref{pb} for the real and imaginary part of $\overline{z_0-z}(t)$, respectively, we finally obtain the average position and momentum of the driven oscillator, $\bt(t)\equiv(\bt_q(t)+i\bt_p(t))/\sqrt{2}$,
\begin{subequations}
\begin{align}   \label{q0-qb}
   \<x(t)\>_{\bt(t)}&=\bt_q(t)=\bar{q}_0(t) 
   +q\left(1-e^{-\gam t/2} \cos(\w t)\right) 
   +e^{-\gam t/2} \frac{\sin(\w t)}{2\w} 
        \big[\th_2 q-(2\w_0 -\th_1)p\big]\,,\\
   \<p(t)\>_{\bt(t)}&=\bt_p(t)=\bar{p}_0(t) 
   +p\left(1-e^{-\gam t/2}\cos(\w t)\right)
     +e^{-\gam t/2} \frac{\sin(\w t)}{2\w} 
        \big[(2\w_0 +\th_1)q-\th_2 p\big]\,,   \label{p0-pb}
\end{align}
\end{subequations}
where $q, p$ are given by \Eqs{qalq}, \eqref{palp}, respectively, in which $\al_q, \al_p$ \eqref{alq}-\eqref{alq} are the coefficients of the linear driving term in $\L_z$ \eqref{Kz}, and $\bar{q}_0(t), \bar{p}_0(t)$ are the displaced position and momentum of the Gaussian mixed state contributed by the initial displacement.
Since $\beta(t\rightarrow\infty)=(q+ip)/\sqrt{2}=z$, the stationary state of the MME \eqref{drhoD} is a displaced Gaussian $\tau(Q,r,t\rightarrow\infty)=\zeta_z(Q,r,0)$, with average position and momentum $q$ and $p$, respectively.
Their magnitudes are determined by the driven term $D(\al)$ in $\L_z$ \eqref{Kz}.
This reaffirms the fact that the MME with the displaced term \eqref{drhoD} indeed describes the dynamics of driven oscillators in open quantum systems.

\section{Liouvillian exceptional points}
\label{SecEP}

Liouvillian exceptional points are points in the parameter space where the eigenvalues and the eigenvectors of the Liouvillian coalesce \cite{Heiss12}. 
The eigenvalues for the GKSL equation and the generic equation were obtained with various methods for the single-mode  \cite{Briegel93,Tay04,Tay08,Honda10,Tay20}, as well as for the multi-mode systems \cite{McDonald23,Gaidash25}.
The structure of the Liouvillian exceptional points of $\L_0$ was studied in Ref.~\cite{Tay23,Gaidash25}.
The eigenvalues of $\L_0$ \cite{Tay08,Tay20} are
\begin{align}   \label{eigevL0}
    \lam_{mn}^\pm&=\pm in\w-(m-n/2)\gam\,, \qquad 0\leq n\leq m\,,
\end{align}
where $\w$ is the renormalized frequency \eqref{w}. 
Exceptional points occur when $\w=0$, i.e., whenever $\th_1$ and $\th_2$ satisfy
\begin{align}   \label{EPth}
    \th_1^2+\th_2^2=4\w_0^2\,,
\end{align}
cf.~\Eq{w}.
The eigenvalues then coalesce to $\lam_\text{E}=-(m-n/2)\gam$ with $n$-fold degeneracies.

From \Eq{EPth}, we learn that the diffusion coefficients $\eta_i$ do not affect the exceptional points. They are determined solely by the coefficients in $H_0$ \eqref{H0}-\eqref{xi}.
For a fixed natural frequency $\w_0$ of the oscillator, the exceptional points can be approached in two independent ways \cite{Tay23}, i.e., when either $\th_1$ or $\th_2$ is zero, while the other one approaches $2\w_0$.
The two independent ways are exemplified by the the Caldeira-Leggett (CL) equation and the modified GKSL equation \cite{Tay23}, respectively.
Referring to \Eq{Kgen}, the CL equation has the parameters $\th_1=0, \th_2=-\gam, \eta_0=\eta_1=-\gam(2\bar{n}+1)$ and $\eta_2=0$. It describes a harmonic oscillator in a high temperature bath, i.e., when the occupation number $\bar{n}$ in the bath mode is large. 

The second way of approaching the exceptional points can be considered by starting with the Liouvillian of the GKSL equation with $\th_1=0=\th_2, \eta_0=-\gam(2\bar{n}+1), \eta_1=0=\eta_2$,
\begin{align}   \label{KL}
    \L_\text{GKSL}\rho&=-i[\w_0\adg a,\rho]+\gam(\bar{n}+1)L\rho
            +\gam\bar{n}R\rho \,.
\end{align}
By adding $-\th_1iM_1$ to $\L_\text{GKSL}$, we obtained a modified Liouvillian of the GKSL equation \cite{Tay23}, $\L_\text{mGKSL}=\L_\text{GKSL}-\th_1iM_1$. The additional term \eqref{iM1} modifies the unitary part of the reduced dynamics into
\begin{align}   \label{KLmodL0}
    \L_\text{mGKSL,0}\rho&=-i\left[\w_0\adg a+\frac{\th_1}{4}(aa+\adg\adg),\rho\right]\,.
\end{align}
To clarify its effect on the reduced dynamics, we write $\L_\text{mGKSL,0}$ in the position representation using \Eqs{iL0}-\eqref{L2+}, and return the dimensions to the coordinates, cf.~\Eq{xpDless}, to obtain
\begin{align}   \label{mKLQr}
    \L_\text{mGKSL,0}(Q,r)&= i\left(\w_0-\frac{\th_1}{2}\right) \frac{\hbar}{m\w_0}\frac{\d^2}{\d Q_d\d r_d} -i\left(\w_0+\frac{\th_1}{2}\right) \frac{m\w_0}{\hbar}Q_d r_d\,.
\end{align}
Introducing an effective mass $m_\text{eff}$ and effective frequency $\w_\text{eff}$, we write the modified unitary part as
\begin{align}   \label{KLmod}
    \L_\text{mGKSL,0}(Q,r)&= i\frac{\hbar}{m_\text{eff}}\frac{\d^2}{\d Q_d\d r_d}-i \frac{m_\text{eff}\w^2_\text{eff}}{\hbar}Q_d r_d\,.
\end{align}
Equating the coefficients of the terms in \Eq{KL} and \eqref{KLmod}, we deduce that
\begin{align} \label{effmass}
    m_\text{eff}=\frac{m}{1-\th_1/(2\w_0)}\,.
\end{align}
Furthermore, the effective frequency is
\begin{align} \label{effw}
    \w_\text{eff}=\sqrt{\w^2_0-\frac{\th_1^2}{4}}\,,
\end{align}
which is consistent with the renormalized frequency of oscillator in \Eq{w}.
Hence, the oscillator acquires an effective mass through the parameter $\th_1$.
As the damping rate or the effective mass of the oscillator approaches $2\w_0$ from below, the oscillatory motion starts getting sluggish. Eventually when \Eq{EPth} is fulfilled and the exceptional point reached, critical-damping occurs. 

Since $\L_z$ is related to $\L_0$ by a similarity transformation \eqref{K'eq}, they have the same eigenvalues.
This is implied by the known fact that the driven Liouvillian differs from its non-driven counterpart by a linear term and is related to it via a unitary displacement transformation \cite{Klauder68,GardinerQNoise,Serafini23}.
Therefore, $\L_z$ and $\L_0$ have the same exceptional points structure.
For $\L_z$, at the exceptional points the average position and momentum are obtained by taking the limit $\w\rightarrow0$ in \Eqs{q0-qb}-\eqref{p0-pb}, respectively, yielding
\begin{subequations}
\begin{align}   \label{xbtEP}
   q_E(t)&=\bar{q}_0(t)+\left(1-e^{-\gam t/2}\right)q
        +\half t e^{-\gam t/2}\big[\th_2 q-(2\w_0 -\th_1)p\big]\,,\\
   p_E(t)&=\bar{p}_0(t)+(1-e^{-\gam t/2})p
        +\half t e^{-\gam t/2}\big[(2\w_0 +\th_1)q-\th_2 p\big]\,.
        \label{pbtEP}
\end{align}
\end{subequations}
Notice the characteristic polynomial time dependence of the form $t\exp(-\gam t/2)$ at the exceptional points,
which reveals that the dynamics corresponds to a critically damped oscillator \cite{Heiss16,Tay23}.
For completeness, time evolution of the parameters in the Gaussian state, $\mu_E(t), \kappa_E(t)$ and $\nu_E(t)$ at the exceptional points are listed in \ref{AppSol}.

When $\th_1$ or $\th_2$ continues to increase beyond the exceptional points, $\w$ turns imaginary $\w=i|\w|$, the previously oscillating functions of the average position \eqref{q0-qb} and momentum \eqref{p0-pb} then depend on hyperbolic sine and hyperbolic cosine, indicating that the system has moved into the overdamped region.
Two characteristic frequencies emerge, $\half \gam\pm|\w|$.
One of the frequencies gives the factor, $\exp[-(\gam/2+|\w|)t]$, which decays rapidly.
The other slower decaying exponential factor, $\exp[-(\gam/2-|\w|)t]$, characterizes the evolution of the system at its later stage of evolution. Once $|\w|$ exceeds $\gam/2$, the solution becomes unstable. We will not consider the latter situation further.

\section{Time-dependent driving}
\label{SecTmDep}

In this section we consider time-dependent driving. Time-dependent driving on the reduced dynamics of an oscillator can be introduced \cite{Vazquez18} through external time-dependent forces on the oscillator, or by adiabatic modulating the frequency of the oscillator. Our approach can be applied to the former. External driving is described by a linear term in the Hamiltonian, 
\begin{align}   \label{H}
    H&=H_0+\lam(t)\adg+\lam^*(t)a\,,
\end{align}
in which $\lam(t)\equiv\lam_r(t)+i\lam_i(t)$ is a time-dependent complex parameter with real $\lam_r$ and $\lam_i$.
The unitary part of the Liouvillian is now $-i[H,\rho]$. The part involving the external driving can then be written in terms of the displacement operator, $-i[\lam(t)\adg+\lam^*(t)a,\rho]=D(-i\lam(t))\rho$.
Consequently, the quantum master equation that we need to consider is
\begin{align}   \label{Ldriv}
    \frac{\d \tau}{\d t}&=\L_0\tau+D(-i\lam(t))\tau\,,
\end{align}
where we have used the density operator $\tau$ anticipating the fact that the solution is a displaced state.

Let us consider solution of the form $\tau=\exp(z(t))\rho$, where $\rho$ is the solution of non-driven MME \eqref{drho} and $z(t)$ is now a time-dependent displacement. 
As a result, the time evolution of $\tau$ acquires an additional time derivative term on $z$,
\begin{align}   \label{drhoDtm}
     \frac{\d \tau}{\d t}
     =e^{D(z(t))}\frac{\d \rho}{\d t}+ D(\dot{z})\tau
    =e^{D(z(t))}\L_0e^{-D(z(t))}\tau+D(\dot{z})\tau
    =\L_0\tau+D(\al(t)+\dot{z})\tau\,,
\end{align}
where we have made use of \Eqs{K'eq} and \eqref{Kz}, and sum the arguments of the displacement operators.
Comparing \Eq{drhoDtm} with \Eq{Ldriv}, we conclude that
\begin{align}   \label{alt}
   \dot{z}+\al(t)&=-i\lam(t)\,.
\end{align}
It leads to a set of two simultaneous first order differential equations in $q$ and $p$,
\begin{subequations}
\begin{align}   \label{DEqt}
   \dot{q}+\half(\gam+\th_2)q-\half(2\w_0-\th_1)p&=\sqrt{2}\lam_i\,,\\
   \dot{p}+\half(2\w_0+\th_1)q+\half(\gam-\th_2)p&=-\sqrt{2}\lam_r\,. \label{DEut}
\end{align}
\end{subequations}
where we use \Eqs{alq}-\eqref{alp} for $\al$ in which $q,p$ are now time-dependent, but we omit the time-dependence on $q, p$ and $\lam_{i,r}$ for simplicity.
\Eq{alt} also implies that the oscillator considered in Section \ref{SecLdisp} is driven by the constant force $\lam=i\al$.

This set of differential equations can be solved with standard method such as the Laplace transform \cite{Arfken} to yield
\begin{subequations}
\begin{align}   \label{qt}
   q(t)&=\bar{q}_0(t)
   +\sqrt{2}\int_{0}^{t} e^{-\gam s/2}\lam_i(t-s)\left(\cos(\w s)-\frac{\th_2}{2\w}\sin(\w s)\right)ds
   -\frac{\th_0-\th_1}{2\w}\sqrt{2}\int_{0}^{t}e^{-\gam s/2}\lam_r(t-s)\sin(\w s) ds\,,\\
   p(t)&=\bar{p}_0(t)
   -\sqrt{2}\int_{0}^{t} e^{-\gam s/2}\lam_r(t-s) \left(\cos(\w s)+\frac{\th_2}{2\w}\sin(\w s)\right)ds
   -\frac{\th_0+\th_1}{2\w}\sqrt{2} \int_{0}^{t}e^{-\gam s/2} \lam_i(t-s) \sin(\w s) ds\,, \label{ut}
\end{align}
\end{subequations}
where $\bar{q}_0(t)$ and $\bar{p}_0(t)$ are contributions from the initial displacement, cf.~\Eqs{qb}-\eqref{pb}, respectively.
We will discuss the effects of a few commonly used external driving forces below.

\begin{figure}[t]
\centering
\includegraphics[width=2.2in,trim = 8cm 11cm 7cm 12cm
]{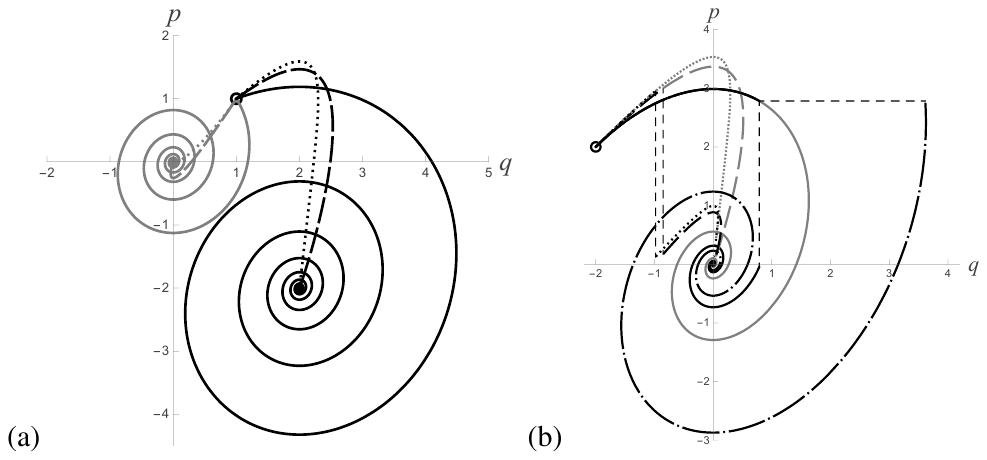}

\caption{
Refer to the main text for the parameters used in the figures. In both figures, gray and black curves denote non-driven and driven oscillator, respectively. Solid curves, dashed curves, and dotted curves label underdamped, critically-damped (at the exceptional point), and overdamped oscillator, respectively. Empty circles indicate initial positions of oscillator. (a) Constant driving. (b) Impulsive driving. When an impulse is introduced, the oscillator is kicked to a new position abruptly. Real impulse with $A=0.5, B=0$ kicks (thin vertical dashed lines) the oscillator vertically at $t=0.5$, imaginary impulse with $B= 0.5, A=0$ kicks (thin horizontal dashed line) the oscillator horizontally at $t=0.5$. The dot-dashed black curve labels underdamped oscillator under imaginary impulse driving.}
\label{fig1}
\end{figure}

\subsection{Constant driving}
\label{ConstDriv}

For constant driving, $\dot{z}=0$, \Eq{alt} gives $\lam_i=\al_q/\sqrt{2}$ and $\lam_r=-\al_p/\sqrt{2}$. We can verify explicitly that \Eqs{qt}-\eqref{ut} correctly yield \Eqs{q0-qb}-\eqref{p0-pb}. For illustrations, we plot the time evolution of the phase space variable for a few cases.
In order to better illustrate the effect of the parameters, we choose the stationary state phase space position $(q,p)$ as the independent variables and obtain the magnitude of the external driving force that generates them through \Eqs{DEqt}-\eqref{DEut}.

In \Fig{fig1}(a), we start with the initial conditions $q_0=1, p_0=1$ (labelled by empty circle) and choose $\gam=1$ as the time scale. Gray curves illustrate the time evolution of initially displaced oscillator without driving $(\bar{q}_0(t), \bar{p}_0(t))$ \eqref{qb}-\eqref{pb}. Black curves depict the phase space motion of oscillator under constant external driving force $(q(t), p(t))$ \eqref{qt}-\eqref{ut}. 
There are three regions of interests.
\begin{enumerate}
\item Underdamped region. With the parameters $\w_0=5, \th_1=1=\th_2$, the oscillator in the underdamped region (solid gray curve) spirals into the origin in the phase space. When a constant driving with $\lam_r=7.778=-\lam_i$ is applied, the oscillator spirals into a new position $q=2, p=-2$ in the phase space, illustrated by solid black curve. The renormalized frequency is $\w=4.950$.
\item Critically-damped region (at the exceptional point). While keeping the other parameters fixed, reducing $\w_0$ to $1/\sqrt{2}\approx0.707$ brings the oscillator to the exceptional point $\w=0$. The motion is represented by the long-dashed gray and long-dashed black curves for a non-driven and driven oscillator, respectively. The external force required to displace the oscillator to $q=2, p=-2$ is $\lam_r=1.707=-\lam_i$. 
\item Overdamped region. Dotted curves give the motion of oscillator in the overdamped region, when $\w_0$ reduces further to $0.6$, causing $\w=0.374 i$ to turn imaginary. Now $|\w|=0.374<\gam/2$. The external force $\lam_r=1.556=-\lam_i$ brings the oscillator to the steady state position at $q=2, p=-2$.
\end{enumerate}
In all the regions, as the restoring force on the oscillator reduces, $\w_0$ decreases. Smaller external forces are then required to displace the oscillator to the same stationary position.

\subsection{Impulsive driving}
\label{ImpulseDriv}

We consider an impulse $\lam(t)=A\del(t-a)+iB\del(t-b)$, where $\del(t)$ is the Dirac-delta function, with the real and imaginary part of the impulse occurring at different instants with different strengths. 
They produce the following motion of the oscillator in the phase space,
\begin{subequations}
\begin{align}   \label{qtimpulse}
   q(t)&=\bar{q}_0(t)
   -\sqrt{2}A e^{-\gam (t-a)/2}(2\w_0-\th_1)\frac{\sin\w(t-a)}{2\w} H(t-a)
   +\sqrt{2}B e^{-\gam (t-b)/2}\left(\cos \w(t-b)-\th_2\frac{\sin\w(t-b)}{2\w}\right) H(t-b)\,,\\
   p(t)&=\bar{p}_0(t)
   -\sqrt{2}A e^{-\gam (t-a)/2}\left(\cos \w(t-a)+\th_2\frac{\sin\w(t-a)}{2\w}\right) H(t-a)
   -\sqrt{2}B e^{-\gam (t-b)/2} (2\w_0+\th_1)\frac{\sin\w(t-b)}{2\w} H(t-b)
   \,, \label{utimpulse}
\end{align}
\end{subequations}
where $H(t)$ is the Heaviside step function.

\Fig{fig1}(b) illustrates the motion of oscillator without external driving (gray curves) and driven by an impulse (black curves). We plot the influence of an impulse on the oscillator with real impulse $A=2$ introduced at $t=a=0.6$, while $B=0$. The black curves show an underdamped oscillator "kicked" (kickings are represented by vertical thin dotted lines) to a new position at $t=0.6$. When the oscillator is kicked to a position closer to the equilibrium by the impulse, it evolves to equilibrium faster than its non-driven counterpart (gray curves). 
The parameters used are $\gam=1, q_0=-2, p_0=2, \th_1=\th_2=1$. For the underdamped (solid curves), critically damped (long-dashed curves) and overdamped (dotted curves) region, we choose $\w_0=2$ giving $\w=1.871$, $\w_0=1/\sqrt{2}$ giving $\w=0$, and $\w_0=0.6$ giving $\w=0.374i$, respectively. 

The impulse can also kick the oscillator further away from the equilibrium, causing it to take a longer time to relax to
equilibrium. 
This is illustrated by the dot-dashed black curve upon imposing an imaginary impulse at the instant $t=b=0.6$, with strength $B=2$ and $A=0$, for the underdamped case when $\w_0=2$. This time, the oscillator is kicked horizontally (thin horizontal dashed line) to a new position further away from the stationary state.
 
The results can be used to understand the influence of constant driving forces of the Heaviside-type, such as $\lam(t)=AH(t-a)+iBH(t-b)$, where the influence is initiated at a latter time and remains constant thereafter. The oscillator will experience a kick at the instant the force is initiated. It then evolves to a displaced equilibrium position as the external driving remains constant, similar to the case of constant driving force.

\begin{figure}[t]
\centering
\includegraphics[width=2.2in,trim = 8cm 10.5cm 7cm 12cm
]{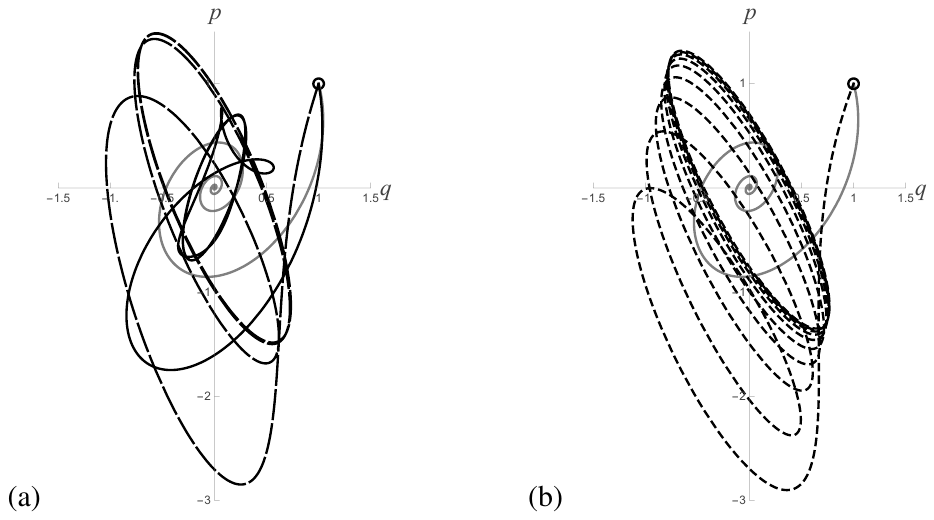}

\caption{Harmonic driving. Refer to the main text for the parameters used in the figures. In both figures, gray and black curves denote non-driven and driven oscillator, respectively. Empty circles at $(1,1)$ denote the initial positions of oscillator. (a) Solid and long-dashed curves label underdamped and critically-damped oscillator (at the exceptional point), respectively. (b) Short-dashed curve labels overdamped oscillator. The oscillator takes a longer time to reach the steady state in the overdamped region.}
\label{fig2}
\end{figure}

\subsection{Harmonic driving}
\label{HarmDriv}

Using an external harmonic driving, $\lam(t)=R \cos\Om t+iR\sin\Om t$, the motion of the average oscillator's position and momentum in the phase space can be cast into the following form, 
\begin{subequations}
\begin{align}   \label{qharmt}
   q(t)&=\bar{q}_0(t) + e^{-\gam t/2} R_\w \cos(\w t-\psi_q) +R_\Om \cos(\Om t-\phi_q)\,,
\end{align}
\begin{align}   \label{uharmt}
   p(t)&=\bar{p}_0(t)   + e^{-\gam t/2}S_{\!\w} \cos(\w t-\psi_p) +S_{\!\Om} \cos(\Om t-\phi_p)\,.
\end{align}
\end{subequations}
They consist of the superposition of three terms.
The first two terms come from the initial displacement \eqref{qb}-\eqref{pb} and oscillatory motion of the reduced dynamics with the renormalized frequency $\w$.
These are exponential decaying terms.
The third terms are caused by the external driving force with frequency $\Om$, to which the oscillator will eventually settle into at the steady state.
Because the expressions of the amplitude of the oscillating terms $R_\Om, R_{\!\w}, S_{\!\Om}, S_\w$ and the phases $\phi_{q,p}, \psi_{q,p}$ are lengthy, we present them in \ref{AppPhase}.

In the underdamped region where the renormalized frequency $\w$ \eqref{w} is real, and in the overdamped region when $\w$ turns imaginary $\w=i|\w|$ but satisfies $|\w|<\gam/2$, after the initial displacement and decaying oscillating terms die down, the motion settles down into a steady state that oscillates at the frequency of the external driving force, 
\begin{align}   \label{ststate}
q_\st= R_\Om\cos(\Om t-\phi_q)\,, \qquad p_\st= S_{\!\Om}\cos(\Om t-\phi_p)\,.
\end{align}
It traces out a rotated ellipse in the phase space
\begin{align}   \label{ellipse}
    \frac{(q_\st\cos\theta+p_\st\sin\theta)^2}{a^2}
    +\frac{(-q_\st\sin\theta+p_\st\cos\theta)^2}{b^2}=1\,,
\end{align}
where the rotation angle $\theta$, the semiminor axis $a$ and the semimajor axis $b$ of the ellipse in terms of the parameters of the Liouvillian are given in \ref{AppPhase}.

In \Fig{fig2}(a) and (b), gray curves depict the motion of the initial displacement with the parameters $\gamma = 1, q_0 = 1, p_0 = 1$.
Black curves give motion of the oscillator under harmonic driving with parameters $\Om= 1, R = 1, \th_1 = 1, \th_2 = 3/2$.
When $\w_0 = 5/2$, the oscillator is in the underdamped region with $\w=2.332$. The trajectory settles down into an ellipse at the steady state, labeled by the solid black curve.
As $\w_0$ reduces to $\sqrt{13}/4\approx0.901$, the oscillator is at the critical region $\w=0$, tracing out the long-dashed curve and finally settles into an ellipse at the steady state.
As $\w_0$ further reduces to $0.8$, $\w\approx 0.415i$ becomes imaginary. The motion is overdamped. Since $|\w|<\gam/2$, the steady state remains an ellipse, illustrated by the short-dashed curve in \Fig{fig2}(b).
Notice that the motion settles down into the ellipse much slower in the overdamped region compared to the underdamped region.

\section{Conclusion}
\label{SecConclusion}

We have studied the generic Markovian quantum Liouvillian for linearly displaced or driven oscillator in open quantum systems and obtain the solutions to the quantum MME in the form of Gaussian states, including the solutions at the exceptional points. 
It is known that the evolution of the displacements or first moments depend solely on the coefficients of the unitary part of the Liouvillian and the system's decay rate, but not on the bath temperature.
We further show that the displacements are also independent of the diffusive coefficients of the fast-rotating modes.
The facts that the supergenerators of the fast-rotating modes annihilate linear operators and that the diffusion coefficients can be transformed through thermal symmetries of the Liouvillian provide plausible explanations for this independence. 
When external time-dependent driving is considered, the results obtained for constant driving remain valid. 
These results, which are based on the generalized form of thermal symmetry, may provide further insight to the study of multimode and $N$-level multipartite systems.

\hfill

\noindent \textbf{Acknowledgments}

\hfill

We thank the anonymous referees for their helpful comments and valuable recommendations of references.

\appendix

\section{Superoperators in various representations}
\label{AppRepres}

\subsection{Superoperators in terms of creation and annihilation operators}
\label{Appa}

The superoperators in $K_0$ \eqref{Kgen} are related to the superoperators $L, R, V$ and $V^\dg$ \eqref{Lrho}-\eqref{Vrho} in the Liouvillian $\L_0$ \eqref{Liouv0} by
\begin{subequations}
\begin{align}   \label{iL0a}
   iL_0\rho&=\frac{i}{2}[\adg a,\rho]\,,\\
   iM_1\rho&=\frac{i}{2}\left[\frac{1}{2}\left(aa+\adg\adg\right),\rho\right]\,,
   &\qquad
   iM_2\rho&=\frac{i}{2}\left[\frac{i}{2}\left(aa-\adg\adg\right),\rho\right]\,,\label{iM1}\\
   (O_0-I/2)\rho&=\frac{1}{2}(R-L)\rho\,,&\qquad
   O_+\rho&=\frac{1}{2}(R+L)\rho\,,\label{O0}\\
   L_{1+}\rho&=\frac{1}{2}(V+V^\dg)\rho\,,&\qquad
   L_{2+}\rho&=\frac{i}{2}(V-V^\dg)\rho\,.\label{L2+a}
\end{align}
\end{subequations}

Below, we clarify the relationship between thermal symmetry \cite{Tay07,Tay17} and dissipative symmetry \cite{McDonald23}.
The symmetry can transform or gauge away the noise in the Liouvillian.
The thermal symmetry \cite{Tay07} and its extension \cite{Tay17} make use of the generators $O_0$ and $O_+, L_{1+}, L_{2+}$, respectively. 
The dissipative symmetry superoperator introduced in Ref.~\cite{McDonald23} is $e^{-(2\bar{n}_\text{th}+1)\aq^\dg\aq}$.
We first rewrite its generator $\aq^\dg\aq$ in our notation, where
the quantum superoperators are $\aq\rho=\frac{
1}{\sqrt{2}}(a\rho-\rho a)$, $\aq^\dg\rho=\frac{
1}{\sqrt{2}}(\adg\rho-\rho\adg)$, and the classical superoperators are $\acl\rho=\frac{
1}{\sqrt{2}}(a\rho+\rho a)$, $\acl^\dg\rho=\frac{
1}{\sqrt{2}}(\adg\rho+\rho\adg)$, see Ref.~\cite{McDonald23}.
Then we find that
\begin{align}
  \aq^\dg\aq\rho=\frac{1}{2}(\adg a\rho-\adg\rho a-a\rho\adg+\rho a\adg)=-\frac{1}{2}(R+L)\rho=-O_+\rho\,,
\end{align} 
using the second equation of \eqref{O0}, i.e., 
the generator of the dissipative symmetry is essentially $O_+=-\aq^\dg\aq$, which is one of the generators of the extended symmetry in \cite{Tay17}.
On the other hand, the generator of the thermal symmetry \cite{Tay07} is
\begin{align}
  O_0\rho=\frac{1}{2}(R-L+I)\rho=\frac{1}{2}(\adg\rho a-a\rho\adg)=\frac{1}{2}(\aq^\dg\acl-\acl^\dg\aq)\rho\,,
\end{align} 
which involves a mixture of quantum and classical superoperators. 
To our knowledge, such a mixture was not considered in \cite{McDonald23}.
Likewise, we can show that the other generators of general symmetry associated with the fast-rotating modes are
\begin{align}
  L_{1+}=-\frac{1}{2}(\aq^{\dg 2}+\aq^2)\,,
  \qquad L_{2+}=\frac{i}{2}(\aq^{\dg 2}-\aq^2)\,.
\end{align} 

\subsection{Superoperators in the coordinate representation}
\label{Appx}

In this section we recall known results from the generic quantum Markovian master equations (MMEs) for harmonic oscillator quadratic in the position coordinates \cite{Tay17b}. 
We shall first introduce the notations we use in the coordinate representation. 
The trace of a density operator $\rho$ is
\begin{align}   \label{trrhoxx}
   \tr(\rho)&\equiv\Iii dx\,\<x|\rho|x\>
   =\Iii dx\Iii dy\, \<x|\rho|y\> \del(x-y)\,,
\end{align}
in which the off-diagonal matrix elements are introduced in the second equality.
In the center and relative coordinates \eqref{Qrxx}, the trace becomes
\begin{align}   \label{trrhoQr}
   \tr(\rho)=\Iii dQ \Iii dr \,\rho(Q,r) \del(r)\,,
\end{align}
where $\rho(Q,r)$ is defined by \Eq{rhoxy}.
The average or trace of an superoperator $O$ acting on the density operator is defined by
\begin{align}   \label{Orho}
   \<O\>=\tr\big( O\rho\big)\equiv\Iii dQ \Iii dr \, \del(r)\,O(Q,r)\rho(Q,r) \,.
\end{align}

\subsection{Pure states}
\label{AppPureSt}

We use dimensionless position $x$ and dimensionless momentum $p$ \eqref{xpDless}.
In the position representation, the annihilation and creation operators are, respectively,
\begin{align}   \label{aaxp}
   a&=\frac{1}{\sqrt{2}}(\hat{x}+i\hat{p})
   =\frac{1}{\sqrt{2}}\left(x+\frac{\d}{\d x}\right)\,, \qquad  \adg=\frac{1}{\sqrt{2}}(\hat{x}-i\hat{p})
   =\frac{1}{\sqrt{2}}\left(x-\frac{\d}{\d x}\right)\,.
\end{align}
We then have
\begin{align}   \label{arho}
   \<x|a|\phi\>=\frac{1}{\sqrt{2}}\left(x+\frac{\d}{\d x}\right)\<x|\phi\>\,, \qquad 
   \<x|\adg|\phi\>=\frac{1}{\sqrt{2}}\left(x-\frac{\d}{\d x}\right)\<x|\phi\>\,.
\end{align}
Taking the complex conjugate, and using $\<x|\phi\>^*=\<\phi|x\>$ and $\<x|a|\phi\>^*=\<\phi|\adg|x\>$, we obtain
\begin{align}   \label{rhoadg}
   \<\phi|a|y\>=\frac{1}{\sqrt{2}}\left(y-\frac{\d}{\d y}\right)\<\phi|y\> \,, \qquad
   \<\phi|\adg|y\>=\frac{1}{\sqrt{2}}\left(y+\frac{\d}{\d y}\right)\<\phi|y\> \,,
\end{align}
after relabelling $x$ by $y$. Further examples are
\begin{align}   \label{aadgphi}
   \<x|a\adg|\phi\>=\frac{1}{\sqrt{2}}\left(x+\frac{\d}{\d x}\right)\<x|\adg|\phi\>
   =\frac{1}{2}\left(x+\frac{\d}{\d x}\right)\left(x-\frac{\d}{\d x}\right)\<x|\phi\>\,,
\end{align}
\begin{align}   \label{phiadga}
   \<\phi|\adg a|y\>=\frac{1}{\sqrt{2}}\left(y-\frac{\d}{\d y}\right)\<\phi|\adg|y\>
   =\frac{1}{2}\left(y-\frac{\d}{\d y}\right)\left(y+\frac{\d}{\d y}\right)\<\phi|y\> \,.
\end{align}

\subsection{Mixed states}
\label{AppMixSt}

Next, we consider mixed states 
\begin{align}   \label{rhophi}
   \rho=\sum_{ij} c_{ij}|\phi_i\>\<\phi_j|\,,
\end{align}
where $\phi_i$ are the eigenstates of free oscillator, i.e., the number states. Then we have, for instance,
\begin{align}   \label{arhoxy}
    \<x|a\rho|y\>=\sum_{ij} c_{ij}\frac{1}{\sqrt{2}}\left(x+\frac{\d}{\d x}\right)\<x|\phi_i\>\<\phi_j|y\>=\frac{1}{\sqrt{2}}\left(x+\frac{\d}{\d x}\right)\<x|\rho|y\>\,,
\end{align}
and so on for other operators. In the $Q$ and $r$ representation, the displacement superoperator \eqref{D} is
\begin{align}   \label{DQr}
   \<x|D(z)\rho|y\>&=\big(z\<x|\adg\rho|y\> -z^*\big(\<x|a\rho|y\>\big)-\big(z\<x|\rho \adg)|y\>-z^*\<x|\rho a|y\>\big)\nn
   &=\frac{1}{\sqrt{2}} \left[ z\left(x-\frac{\d}{\d x}\right)
   -z^*\left(x+\frac{\d}{\d x}\right)
   -z\left(y+\frac{\d}{\d y}\right) +z^*\left(y-\frac{\d}{\d y}\right)
   \right]\<x|\rho|y\>\nn
   &=\frac{1}{\sqrt{2}} \left[ z\left(r-\frac{\d}{\d Q}\right)-z^*\left(r+\frac{\d}{\d Q}\right) \right] \left\<Q+\frac{r}{2}\bigg|\rho\bigg|Q-\frac{r}{2}\right\> \nn
   &=\left(ipr-q\frac{\d}{\d Q}\right) \rho(Q,r)\,,
\end{align}
which leads to \Eq{Dx}.

In this way we obtain the position representation of the superoperators in $K_0$ \eqref{iL0a}-\eqref{L2+a},
\begin{subequations}
\begin{align} \label{iL0}
    iL_0 &= \frac{i}{2}\left(-\frac{\d^2}{\d Q\d r}+Qr\right)\,,\qquad
    iM_1= \frac{i}{2}\left(\frac{\d^2}{\d Q\d r}+Qr\right)\,,\qquad
    iM_2= -\frac{1}{2}\left(\frac{\d}{\d Q}Q+r\frac{\d}{\d r}\right)\,,\\
    O_0 &= -\frac{1}{2}\left(Q\frac{\d}{\d Q}-r\frac{\d}{\d r}\right)\,,\qquad
    O_+ = \frac{1}{4}\left(\frac{\d^2}{\d Q^2}-r^2\right)\,, \qquad
    L_{1+} = -\frac{1}{4}\left(\frac{\d^2}{\d Q^2}+r^2\right)\,,\qquad
    L_{2+} = -\frac{i}{2}r\frac{\d}{\d Q}\,. \label{L2+}
\end{align}
\end{subequations}
The coordinate representation of the superoperators in the Liouvillian $\L_0$ can be obtained by linear combinations of \eqref{iL0}-\eqref{L2+} using \eqref{iL0a}-\eqref{L2+a}, respectively.

We next discuss the requirements that should be imposed on the density operator so that the cyclic permutations of operators inside traces are valid.
We use $\tr(a\rho)$ as an example. Using the definition of trace in \Eqs{trrhoxx} and \eqref{rhophi} for mixed states, we obtain 
\begin{align}   \label{tra}
   \tr(a\rho)&=\sum_{ij} c_{ij} \Iii dx\,\left[\left(x+\frac{\d}{\d x}\right)\<x|\phi_i\>\right]\<\phi_j|x\>\nn
   &=\sum_{ij} c_{ij}\left[ \Iii dx\,\<x|\phi_i\>\left(x-\frac{\d}{\d x}\right)\<\phi_j|x\>+ \<x|\phi_i\>\<\phi_j|x\>\bigg|_{-\infty}^\infty\right]\nn
   &=\tr(\rho a)\,,
\end{align}
where in the second equality we have carried out an integration by parts to yield a boundary term. 
For the cyclic permutation under the trace to be valid to yield the last equality, we require the eigenfunctions $\<x|\phi_i\>$, or more generally $\<x|\rho|y\>$, to be (i) differentiable, and (ii) vanish fast enough at infinity in the $x$ and $y$ coordinates. This is true for the eigenfunctions of free harmonic oscillator and the Gaussian mixed states $\zeta_0$ \eqref{rhog} considered here.

\subsection{Generic form of displacement superoperator}
\label{AppDform}

In this appendix we impose two requirements on a generic superoperator \cite{Prigogine73,Tay17} , namely, (i) adjoint-symmetry and (ii) tracelessness with $\rho$, to obtain the generic form of displacement superoperator \eqref{D} in the quantum Liouville space.
We start with a generic form of superoperator linear in the creation and annihilation operators,
\begin{align}   \label{G}
   G&\equiv s(a\times 1)+t(\adg\times 1)+u(1\times \adg)+v(1\times a)\,,
\end{align}
where $s, t, u, v$ are complex numbers.
For arbitrary operators $a, b$ and complex number $s$, $s(a\times b)$ acts on $\rho$ as $s(a\times b)\rho=sa\rho b$ \cite{Prigogine73,Tay17}.  
The association operation \cite{Prigogine73,Tay17} is defined by 
\begin{align}   \label{association}
\widetilde{s(a\times b)}=s^*b^\dg\times a^\dg \,,
\end{align}
leading to
\begin{align}   \label{Gt}
   \tilde{G}&\equiv s^*(1\times \adg) +t^*(1\times 1) +u^*(a\times1 ) +v^*(\adg\times 1)\,.
\end{align}
From the adjoint-symmetric requirement, $\tilde{G}=G$, we deduce that $u=s^*$ and $v=t^*$ to obtain
\begin{align}   \label{Gadj}
   G&= s(a\times 1)+t(\adg\times 1)+s^*(1\times \adg)+t^*(1\times a)\,.
\end{align}
To preserve the trace of density matrix, a generator $G$ of an equation of motion, for example, $\d \rho/\d t=G\rho$, should have zero trace on the density matrix, $\tr(G\rho)=0$, so that $\partial(\tr\rho)/\partial t=\tr(G\rho)=0$, giving constant $\tr\rho=1$.
Using cyclic permutation of operators under the trace we obtain
\begin{align}   \label{Gadjtr}
   \tr(G\rho)&=\tr( sa\rho+t\adg\rho+s^*\adg\rho+t^*a\rho)\nn
   0&=(s+t^*)\<a\rho\>+(t+s^*)\<\adg\rho\>
\end{align}
This can always be satisfied provided $t=-s^*$ for arbitrary density operator.
Consequently, we obtain a superoperator linear in the creation and annihilation operators,
\begin{align}   \label{Gs}
   G(s)&= s(a\times 1-1\times a)-s^*(\adg\times 1-1\times \adg)\,.
\end{align}
Upon setting $s=-z^*$, we obtain the generator of the displacement superoperator $D(z)$ \eqref{D}.
We can also choose $s=(i\lam)^*$, so that
$G(-i\lam^*)=-i\left[(\lam^* a+\lam \adg)\times 1-1\times (\lam^* a+\lam \adg)\right]$.
This is a common way of representing the effect of external driving fields on the system \cite{GardinerQNoise}, compare with \Eq{H}.

\section{Equations of motion for first moments}
\label{App1stmoment}

In this appendix, we obtain the equations of motion for the first moments of a quantum oscillator satisfying the generic MME in \Eqs{drho} and \eqref{Liouv0}.
We will show explicitly that the bath temperature included in $\eta_0$ and the parameters for the fast-rotating modes $\eta_1, \eta_2$ will not affect the time evolution of the first moments. To our knowledge, this conclusion has previously been obtained for the GKSL equation \cite{Ma18,Serafini23} without including fast-rotating modes, which are retained in the present work.

To calculate the equations of motion for the first moments, we need to evaluate expressions such as
\begin{align}   \label{1stmomeq}
        \frac{d}{dt}\<x\>&=\tr\left(\hat{x}\frac{\d\rho}{\d t}\right)=\tr(\hat{x}\L_0\rho)= \tr(\L_0^\T(\hat{x})\rho)\,,
\end{align}
and a similar expression for the momentum operator.
In \Eq{1stmomeq}, we have introduced the transposition operation on superoperator \cite{Prigogine73,Tay17,Tay19b} denoted by $\T$.
For a superoperator $S=\sum_ic_i A_i\times B_i$ with the action $S\rho=\sum_ic_i A_i\rho B_i$, where $c_i$ is a complex coefficient, its transposition is
\begin{align}
        S^\T=\sum_ic_i B_i\times A_i\,.
\end{align}
Using \Eq{1stmomeq}, the transpositions of the seven superoperators listed in \Eqs{iL0a}-\eqref{L2+} are \cite{Tay19b}
\begin{subequations}
\begin{align}
        (iL_0)^\T&=-iL_0\,, \qquad (iM_1)^\T=-iM_1\,, \qquad 
        (iM_2)^\T=-iM_2\,, \qquad (O_0-I/2)^\T=-(O+I/2)\,,\\
        O_+^\T&=O_+\,, \qquad\qquad L_{1+}^\T=L_{1+}\,, \qquad\qquad L_{2+}^\T=L_{2+}\,.
\end{align}
\end{subequations}
We can now verify explicitly that $O^\T_+, L^\T_{1+}$ and $L^\T_{2+}$ annihilate linear operators $a, \adg$ and $\hat{x}, \hat{p}$, i.e., $O^\T_+(\hat{x})=0$, and etc.
Therefore, $\eta_0, \eta_1, \eta_2$ drop out from the equation of motion for the first moments \eqref{1stmomeq}.

We are then left with the supergenerators of unitary part and $O-I/2$ in $\L_0^\T$. 
Acting them on $\hat{x}, \hat{p}$ and tracing the results over with $\rho$, we obtain the following equations of motion,
\begin{align}
    \frac{d}{dt}\<x\>=-\frac{1}{2}(\gam+\th_2)\<x\>+\frac{1}{2}(2\w_0-\th_1)\<p\>\,,
    \qquad
    \frac{d}{dt}\<p\>=-\frac{1}{2}(2\w_0+\th_1)\<x\>-\frac{1}{2}(\gam-\th_2)\<p\>\,.
\end{align}
They are consistent with the left-hand-side of the equations for the first moments under time-dependent external driving \eqref{DEqt}-\eqref{DEut}.

\section{Non-hermitian eigenvalue problem}
\label{AppEigevProb}

In this appendix, we calculate the displacement parameter $\bar{z}$ defined in \Eq{KDK}. We first list a few useful commutation relations,
\begin{align}   \label{L0comm}
   [iL_0, D(z)]&=\frac{1}{2}D(iz)\,,\qquad
   [iM_1, D(z)]=-\frac{1}{2}D\big((iz)^*\big)\,,\\
      [iM_2, D(z)]&=\frac{1}{2}D(z^*)\,,\qquad
   [O_0-I/2, D(z)]=\frac{1}{2}D(z)\,.\label{O0comm}
\end{align}
Three other superoperators, $O_+, L_{1+}$ and $L_{2+}$, separately commute with $D(z)$.
Using the commutation relations, we obtain the first equation in \Eq{commDK}, where $\al$ is given by \Eqs{alz}-\eqref{alp}.
We can then use the results to calculate the parameter $\zb$ defined in \Eq{KDK},
\begin{align}   \label{Dzb}
    D(\zb)&\equiv e^{\L_0t}D(z)e^{-\L_0t}=D(z)+t[\L_0,D(z)]
    +\frac{t^2}{2!}\left[\L_0,[\L_0,D(z)]\right]+\cdots\,.
\end{align}
We can simplify the calculation by casting \Eq{Dzb} into a matrix form. 
We write $D(z)$ \eqref{D} as a column vector
\begin{align}   \label{zvec}
    \DB{z}&=\left(\begin{array}{c}
         z \\
         -z^*
       \end{array}\right)\,.
\end{align}
The commutator $[\L_0,D(z)]$ can then be written as a matrix multiplication. The first commutation relation in \Eq{commDK} can be represented by
\begin{align}   \label{KqBDzb}
    \LqB\cdot\DB{z}=-\DB{\al}\,,
\end{align}
where $\LqB$ is a $2\times2$ matrix,
\begin{align}   \label{KqB}
    \LqB&\equiv-\left(\begin{array}{cc}
         a & b \\
         b^* & a^*
       \end{array}\right)\,,
\end{align}
with matrix elements
\begin{align}   \label{a}
    a\equiv\half\gam+i\w_0\,,\qquad
    b\equiv -\half(\th_2+i\th_1)\,.
\end{align}
Then,
\begin{align}
    \al=\frac{1}{\sqrt{2}}(\al_q+i\al_p)=az-bz^*\,,
\end{align}
where the explicit expressions of $\al_q$ and $\al_p$ are given in \Eqs{alq} and \eqref{alp}, respectively.
\Eq{Dzb} can then be further written as
\begin{align}   \label{DBzb}
    \DB{\zb}&=\DB{z}+t\LqB\cdot\DB{z}+\frac{t^2}{2!}(\LqB)^2\cdot\DB{z}+\cdots
    =e^{t\LqB}\cdot\DB{z}\,.
\end{align}
To calculate the right-hand-side (RHS) of \Eq{DBzb}, we need to consider the right and left eigenvalue problem of the non-Hermitian matrix $\LqB$,
\begin{align}   \label{Kqv}
    \LqB\cdot\vB_\pm=-\lam_\pm\vB_\pm\,,\qquad
    \boldsymbol{L}_0^\dg\cdot\uB_\pm=-\lam^*_\pm\uB_\pm\,,
\end{align}
respectively, where $\lam_\pm$ are complex eigenvalues,
\begin{align}  \label{lam}
    \lam_\pm&\equiv \half\gam\pm i\w\,,
\end{align}
in which the renormalized frequency $\w$ is defined in \Eq{w}.
The right eigenvector and left eigenvector are, respectively,
\begin{align}   \label{vBpm}
    \vB_\pm&=N_\pm
            \left(
              \begin{array}{c}
                \lam_\pm-a^* \\
                b^* \\
              \end{array}
            \right)\,,\qquad
     \uB_\pm=N^*_\pm
            \left(
              \begin{array}{c}
                \lam^*_\pm-a \\
                b^* \\
              \end{array}
            \right)\,,
\end{align}
where $N_\pm$ are normalization constants,
\begin{align}   \label{N}
    N_\pm&=\frac{1}{\sqrt{(\lam_\pm-a^*)^2+|b|^2}}=\frac{-i}{\sqrt{2\w(\w\pm\w_0)}}\,.
\end{align}
The eigenvectors form a biorthogonal set, $\uB_\pm^\dg\cdot\vB_\pm=1$ and $\uB_\pm^\dg\cdot\vB_\mp=0$.
We can then diagonalize $\LqB$ through
\begin{align}   \label{UKqV}
    \UB^\dg\cdot\LqB\cdot\VB=-\left(
                               \begin{array}{cc}
                                 \lam_+ & 0 \\
                                 0 & \lam_- \\
                               \end{array}
                             \right)    \,,
\end{align}
where
\begin{align}   \label{VB}
    \VB&=\left(
          \begin{array}{cc}
            \vB_+ & \vB_- \\
          \end{array}
        \right)
        =\left(
           \begin{array}{cc}
             N_+(\lam_+-a^*) & N_-(\lam_--a^*) \\
             N_+b^* & N_-b^* \\
           \end{array}
         \right)\,,\\
    \UB&=\left(
          \begin{array}{cc}
            \uB_+ & \uB_- \\
          \end{array}
        \right)
        =\left(
           \begin{array}{cc}
             N_+^*(\lam_+^*-a) & N_-^*(\lam_-^*-a) \\
             N_+^*b^* & N_-^*b^* \\
           \end{array}
         \right)\,.
\end{align}
We can show that they satisfy
\begin{align}   \label{VU1}
    \UB^\dg\cdot\VB=\boldsymbol{I}\,, \qquad \VB\cdot\UB^\dg=\boldsymbol{I}\,.
\end{align}
Inserting $\VB\cdot\UB^\dg=\boldsymbol{I}$ into the RHS of \Eq{DBzb}, we obtain
\begin{align}   \label{DBzbsol}
    \DB{\zb}&=e^{\LqB t}\cdot\DB{z}\nn
    &=\VB\cdot\UB^\dg\cdot e^{\LqB t}\cdot\VB\cdot\UB^\dg\cdot\DB{z}\nn
    &=\VB\cdot e^{\UB^\dg\cdot\LqB\cdot\VB t}\cdot\UB^\dg\cdot\DB{z} \nn
    &=\VB\cdot\left(
               \begin{array}{cc}
                 \exp(-\lam_+ t) & 0 \\
                 0 & \exp(-\lam_- t) \\
               \end{array}
             \right)\cdot\UB^\dg\cdot\DB{z}\,,
\end{align}
where we use \Eq{UKqV} to diagonalize $\exp(\UB^\dg\cdot\LqB\cdot\VB t)$. 
We can now multiply the matrices in the last line of \Eq{DBzbsol}, and equate the results to $\DB{\zb}=(\zb, -\zb^*)^T$.
After simplifying the expressions, we obtain
\begin{align}   \label{zb}
    \zb(t)&=\frac{1}{4\w}e^{-\lam_+ t}\left(\left(2\w+\th_0\right)z
    +\left(\th_1-i\th_2\right)z^*\right)
    +\frac{1}{4\w}e^{-\lam_- t}\left(\left(2\w-\th_0\right)z
    -\left(\th_1-i\th_2\right)z^*\right)\,.
\end{align}
\Eq{zb} leads to the displaced position $\bar{q}$ and displaced momentum $\bar{p}$ in \Eqs{qb} and \eqref{pb}, respectively.

\section{Solutions to quantum MME for Gaussian mixed states}
\label{AppSol}

In this appendix we summarise the solutions to the quantum MME \eqref{drho} for Gaussian mixed states \eqref{rhog} obtained in Ref.~\cite{Tay17b}. 
The parameters in the generic Gaussian mixed states in \Eq{rhog} have the solutions
\begin{align}   \label{mut}
    \mu(t)&=\frac{1}{R(t)}\mu_0 \,,\\
   \nu(t)&=\frac{1}{R(t)}\left( (\mu_0+\nu_0)e^{-2\gam t}
   -\mu_0\big(1+\gv^2\big)
        + e^{-\gam t}\left[\Phi(\gv)\cos(2\w t)
        +\Phi(\thv\wedge\gv)\frac{\sin(2\w t)}{2\w}
        -\thv\cdot\gv \Phi(\thv)\frac{1-\cos(2\w t)}{4\w^2}\right]
          \right)\,,\\
    \kap(t)&=\frac{1}{R(t)}\left(-2\mu_0 g_2
            +e^{-\gam t}\left[\kap_0\cos(2\w t)
            +\half\left[(1-\Del_0^2)\th_0+(1+\Del_0^2)\th_1\right]\frac{\sin(2\w t)}{2\w}
            -\th_2\Phi(\thv)\frac{1-\cos(2\w t)}{4\w^2}\right]
            \right)\,,\\
    R(t)&=2\mu_0[g_0-g_1]+e^{-\gam t}\left[ \cos(2\w t)
    -[\th_2+\kap_0(\th_0-\th_1)]\frac{\sin(2\w t)}{2\w}
    +(\th_0-\th_1)\Phi(\thv)\frac{1-\cos(2\w t)}{4\w^2}\right]\,,\\    \gv&=\frac{1}{\gam^2+4\w^2}\left\{
    \etv\left[-\gam+e^{-\gam t}\big(\gam\cos(2\w t)-2\w \sin(2\w t)\big)\right]
    +\thv\wedge\etv\left[1- e^{-\gam t} \left(\cos(2\w t) +\gam\frac{\sin(2\w t)}{2\w} \right)\right]\right.\nn
    &\quad\left.(\etv\cdot\thv)\frac{\thv}{\gam} \left[1-e^{-\gam t} \left(1+\gam\frac{\sin(2\w t)}{2\w}+\gam^2\frac{1-\cos(2\w t)}{4\w^2} \right)\right]
    \right\}\,,\label{gt} \\
    \Phi(\thv)&=\half(1+\Del^2_0)\th_0+\half(1-\Del^2_0)\th_1+\kap_0 \th_2\,.
\end{align}
In these equations, we have introduced three-component vectors such as $\thv=\th_0\ev_0+\th_1\ev_1+\th_2\ev_2$, and similarly for $\gv$ and $\etv$. The basis vectors $\ev_i$ are orthogonal to each others. 
They have the dot products $\ev_0\cdot\ev_0=-1$, $\ev_1\cdot\ev_1=1$, $\ev_2\cdot\ev_2=1$, and $\ev_i\cdot\ev_j=0$ for $i\neq j$ \cite{Tay17b}. 
The wedge product is defined by $\ev_0\wedge\ev_1=-\ev_2$, $\ev_1\wedge\ev_2=\ev_0$ and $\ev_2\wedge\ev_0=-\ev_1$.

As already discussed in Ref.~\cite{Tay17b}, stable solutions exist in the overdamped region when $\w=i|\w|$ is imaginary but satisfies $|\w|<\gam/2$,
\begin{align}   \label{st}
   \mu_\st&=\frac{1}{2(\Gam_0-\Gam_1)}\,,\qquad \nu_\st=\frac{-\Gv^2-1}{2(\Gam_0-\Gam_1)} \,, \qquad
   \kap_\st=\frac{-\Gam_2}{\Gam_0-\Gam_1}\,,
\end{align}
\begin{align}   \label{Gam}
   \Gv &=\frac{1}{\gam^2+4\w^2}\left(-\gam \etv
   +(\etv\cdot\thv)\frac{\thv}{\gam}+\thv\wedge\etv\right)\,.
\end{align}
For $|\w|\geq\gam/2$, $\mu(t)$ behaves like $e^{(2|\w|-\gam)t}$ or $e^0=1$ in the limit $t\rightarrow\infty$, respectively.
This results in non-normalizable Gaussian mixed states and indicates that stable solution does not exist.

At the Liouvillian exceptional points \eqref{EPth}, by taking the limit $\w\rightarrow0$ on \Eqs{mut}-\eqref{gt}, the solutions of the parameters are
\begin{align}   \label{mucrit}
    \mu_E(t)&=\frac{\mu_0}{R_E}\,,\\
    \nu_E(t)&=\frac{1}{R_E}\left[-\mu_0(1+\gv^2)
    +e^{-\gam t}\left(\Phi(\gv)+\Phi(\thv\wedge\gv)t -\half(\thv\cdot\gv)\Phi(\thv)t^2\right)
    +e^{-2\gam t}(\mu_0+\nu_0)\right]\,,\\
    \kap_E(t)&=\frac{1}{R_E}\left[-2\mu_0g_2+ e^{-\gam t} \left(
    \kap_0+\half\big[(1-\Del_0^2)\th_0+(1+\Del_0^2)\th_1\big]t
    -\half\th_2\Phi(\thv)t^2\right)
    \right]\,,\\ 
    \label{RE}
    R_E(t)&=2\mu_0(g_0-g_1)+e^{-\gam t}\left(1-\big[\th_2+\kap_0(\th_0-\th_1)\big]t+\half(\th_0-\th_1)\Phi(\thv)t^2\right)\,,\\
    \gv_E(t)&=-\frac{\etv}{\gam}(1-e^{-\gam t})
    +\frac{\thv\wedge\etv}{\gam^2}\left[1-e^{-\gam t}(1+\gam t)\right]
    +\frac{(\thv\cdot\etv)\thv}{\gam^3}\left[1-e^{-\gam t}\left(1+\gam t+\half \gam^2t^2\right)\right]
    \,.\label{g}
\end{align}

\section{Expressions for harmonic driving}
\label{AppPhase}


In this appendix, we give the expressions of the phase space motion of the oscillator under harmonic driving in Sec.~\ref{HarmDriv}. 
\Eqs{qt}-\eqref{ut} can be integrated over the harmonic driving $\lam(t)=R \cos\Om t+iR\sin\Om t$. 
The resulting expressions are \Eqs{qharmt}-\eqref{uharmt}, where
\begin{subequations}
\begin{align}   \label{Romcq}
   R_\Om \cos\phi_q\equiv  c_{\!q}(\th_0,\th_1\th_2)
   =\frac{-R}{N\sqrt{2}}
   \left[ (\th_0-\th_1)\left(\frac{\gam^2}{4}+\w^2-\Om^2\right)-\gam\Om\th_2
   +2\Om\left(\frac{\gam^2}{4}-\w^2+\Om^2\right)\right]\,,
\end{align}
\begin{align}   \label{Romsq}
    R_\Om \sin\phi_q\equiv s_{\! q}(\th_0,\th_1\th_2) =\frac{-R}{N\sqrt{2}}
   \left[ \th_2\left(\frac{\gam^2}{4}+\w^2-\Om^2\right)
   +\gam\Om (\th_0-\th_1)-\gam\left(\frac{\gam^2}{4}+\w^2+\Om^2\right)\right]\,,
\end{align}
\begin{align}   
   R_\w \cos\psi_q
   =-c_q(\th_0,\th_1,\th_2)\,,
\end{align}
\begin{align}   
    R_\w \sin\psi_q
    =\frac{R}{\w N \sqrt{2}}
   \left[ \frac{\gam}{2}(\th_0-\th_1)\left(\frac{\gam^2}{4}+\w^2+\Om^2\right)
   -2\gam\Om\w^2 -\Om \th_2 \left(\frac{\gam^2}{4}-\w^2+\Om^2\right)\right]\,.
\end{align}
\end{subequations}
\begin{subequations}
\begin{align}
  S_{\!\Om}\cos\phi_p=-s_q(\th_0,-\th_1,-\th_2)\,,
\end{align}
\begin{align}
  S_{\!\Om}\sin\phi_p= c_q(\th_0,-\th_1,-\th_2)\,,
\end{align}
\begin{align}
  S_{\!\w}\cos\psi_p= s_q(\th_0,-\th_1,-\th_2)\,,
\end{align}
\begin{align}   \label{Sw}
  S_{\!\w}\sin\psi_p= \frac{R}{\w N \sqrt{2}}
   \left[ -2\w^2\left(\frac{\gam^2}{4}+\w^2-\Om^2\right) 
   +\frac{\gam}{2}\th_2\left(\frac{\gam^2}{4}+\w^2+\Om^2\right)
   -\Om(\th_0+\th_1)\left(\frac{\gam^2}{4}-\w^2+\Om^2\right) \right]\,.
\end{align}
\end{subequations}
where
\begin{align} 
    N&=\left[\frac{\gam^2}{4}+(\w-\Om)^2\right]
   \left[\frac{\gam^2}{4}+(\w+\Om)^2\right]
\end{align}
The amplitude of the oscillating terms are
\begin{subequations}
\begin{align}   
  R_\Om(\th_0,\th_1,\th_2)&=\frac{R}{\sqrt{2N}}\sqrt{(\gam-\th_2)^2+\big[2\Om-(\th_0-\th_1)\big]^2}\,,
\end{align}
\begin{align}    
  R_{\w}=\frac{R}{\w\sqrt{2N}}
  \sqrt{\w^2\big[2\Om-(\th_0-\th_1)\big]^2 +\left[\Om\th_2-\frac{\gam}{2}(\th_0-\th_1)\right]^2}\,,
\end{align}
\begin{align}   
  S_{\!\Om}(\th_0,\th_1,\th_2)&=R_\Om(\th_0,-\th_1,-\th_2)\,,
\end{align}
\begin{align}   
  S_{\!\w}=\frac{R}{\w\sqrt{2N}}
  \sqrt{\left[\frac{\gam}{2}\th_2-\Om(\th_0+\th_1)\right]^2
  +4\w^2\left[\frac{\gam^2}{4}+\w^2+\frac{\th_2^2}{4}-\Om(\th_0+\th_1)\right]} \,.
\end{align}
\end{subequations}
The phase of the oscillations $\phi_{q,p}$ and $\psi_{q,p}$ can be determined from \Eqs{Romcq}-\eqref{Sw}.

A conic section in the phase space has the form of a quadratic equation,
\begin{align} \label{ellipseeq}
    1=Aq^2+Bqp+Cp^2=\left(\begin{array}{cc}
      q & p 
    \end{array}\right)M
    \left(\begin{array}{c}
      q\\
      p
    \end{array}\right),
\end{align}
where
\begin{align} \label{M}
    M=\left(\begin{array}{cc}
      A & B/2\\
      B/2 & C
    \end{array}\right).
\end{align}
A geometrical figure described by \Eq{ellipseeq} is (1) an ellipse when the determinant of $M$ satisfies $4|M|=4AC-B^2>0$, (2) a parabola when $4|M|=0$, and (3) a hyperbola when $4|M|<0$.
 
Let a rotation
\begin{align} 
    \underline{R}=\left(\begin{array}{cc}
      \cos\theta & -\sin\theta\\
      \sin\theta & \cos\theta
    \end{array}\right)
\end{align}
diagonalizes $M$ through $\underline{R}^TM\underline{R}=\text{diag}(\lam_+ \quad \lam_-)$,
where $\lam_\pm$ are the eigenvalues of $M$,
\begin{align} 
    \lam_\pm&=\frac{1}{2}(A+C)\pm\frac{1}{2}\sqrt{(A-C)^2+B^2}\,.
\end{align}
For an ellipse, the semiminor and semimajor axis are
\begin{align} \label{ab}
    a=\frac{1}{\sqrt{\lam_+}}\,, \qquad b=\frac{1}{\sqrt{\lam_-}}\,, 
\end{align}
respectively.
The ellipse is rotated in the counter-clockwise direction (positive angle) through an angle $\theta$ determined by
\begin{align}   \label{theta}
    \tan2\theta=\frac{B}{A-C}\,.
\end{align}
\Eq{theta} can be obtained by comparing \Eq{ellipseeq} with \Eq{ellipse} to yield
\begin{align}  
    A=\frac{\cos^2\th}{a^2}+\frac{\sin^2\th}{b^2},\qquad
    B=\sin2\th\left(\frac{1}{a^2}-\frac{1}{b^2}\right)\,,\qquad
    C=\frac{\cos^2\th}{b^2}+\frac{\sin^2\th}{a^2}\,.
\end{align}
We then obtain
\begin{align}  
    A-C=\cos2\th\left(\frac{1}{a^2}-\frac{1}{b^2}\right)\,,
\end{align}
and as a result, we obtain \Eq{theta}.

Starting with the steady state motion in \Eq{ststate}, 
\begin{align}   \label{stOm}
    \left(\begin{array}{c}
            q_\st \\
            p_\st 
          \end{array} \right)
    &=U
          \left(\begin{array}{c}
            \cos \Om t \\
            \sin \Om t 
          \end{array} \right)\,,\\
    U&=\left(\begin{array}{cc}
            c_q & s_q \\
            -s'_q & c'_q 
          \end{array} \right)\,,
\end{align}
where for simplicity we omit the dependence of $\th_i$ in $c_q$ \eqref{Romcq} and $s_q$ \eqref{Romsq}, and abbreviate $c_q'\equiv c_q(\th_0,-\th_1,-\th_2)$, where the prime denotes reflection in the sign of $\th_1$ and $\th_2$, and similarly for $s_q'$.
We can invert \Eq{stOm}, and after setting $\cos^2\Om t+\sin^2\Om t=1$, we obtain the quadratic equation for an ellipse \eqref{ellipseeq}, with the coefficients
\begin{align}  
    A=\frac{1}{|U|^2}({c'_q}^2+{s'_q}^2)\,,\qquad
    B=\frac{2}{|U|^2}(c_q s'_q-s_qc'_q)\,,\qquad
    C=\frac{1}{|U|^2}(c_q^2+s_q^2)\,,
\end{align}
where $|U|=c_qc'_q+s_qs'_q$ is the determinant of $U$.
In this way, the semiminor axis $a$ and semimajor axis $b$ of the ellipse, and the rotated angle of the ellipse $\th$, can be obtained from \Eq{ab} and \Eq{theta}, respectively. 
We find that $4|M|=4AC-B^2=4(c_qs'_q+s_qc'_q)^2/|U|^4>0$, cf.~the sentence below \Eq{M}. 
Therefore, the steady states always trace out an ellipse in the phase space.

\providecommand{\noopsort}[1]{}\providecommand{\singleletter}[1]{#1}%


\begin{thebibliography}{10}
\expandafter\ifx\csname url\endcsname\relax
  \def\url#1{\texttt{#1}}\fi
\expandafter\ifx\csname urlprefix\endcsname\relax\def\urlprefix{URL }\fi
\expandafter\ifx\csname href\endcsname\relax
  \def\href#1#2{#2} \def\path#1{#1}\fi

\bibitem{Walls08}
D.~F. Walls, G.~J. Milburn, Quantum Optics, 2nd Edition, Springer, Berlin,
  2008.

\bibitem{Nielsen}
M.~A. Nielsen, I.~L. Chuang, Quantum Computation and Quantum Information,
  Cambridge, New York, 2000.

\bibitem{Adesso14}
G.~Adesso, S.~Ragy, A.~R. Lee,
  \href{https://doi.org/10.1142/S1230161214400010}{Continuous variable quantum
  information: Gaussian states and beyond}, Open Syst. Info. Dyn. 21 (2014)
  1440001.
\newblock \href {https://doi.org/10.1142/S1230161214400010}
  {\path{doi:10.1142/S1230161214400010}}.

\bibitem{Decoh03}
E.~Joos, H.~D. Zeh, C.~Kiefer, D.~J.~W. Giulini, J.~Kupsch, I.~O. Stamatescu,
  Decoherence and the Appearance of a Classical World in Quantum Theory, 2nd
  Edition, Springer, Berlin, 2003.

\bibitem{Simon87}
R.~Simon, E.~C.~G. Sudarshan, N.~Mukunda,
  \href{http://link.aps.org/doi/10.1103/PhysRevA.36.3868}{Gaussian-wigner
  distributions in quantum mechanics and optics}, Phys. Rev. A 36 (1987)
  3868--3880.
\newblock \href {https://doi.org/10.1103/PhysRevA.36.3868}
  {\path{doi:10.1103/PhysRevA.36.3868}}.

\bibitem{GardinerStoch}
C.~W. Gardiner, Stochastic Methods: A Handbook for the Natural and Social
  Sciences, 4th Edition, Springer, Berlin, 2009.

\bibitem{Serafini23}
A.~Serafini, Quantum Continuous Variables: A Primer of Theoretical Methods, 2nd
  Edition, CRC Press, New York, 2023.

\bibitem{Hillery84}
M.~Hillery, R.~O'Connell, M.~Scully, E.~Wigner,
  \href{http://www.sciencedirect.com/science/article/pii/0370157384901601}{Distribution
  functions in physics: Fundamentals}, Phys. Rep. 106 (1984) 121--167.
\newblock \href {https://doi.org/https://doi.org/10.1016/0370-1573(84)90160-1}
  {\path{doi:https://doi.org/10.1016/0370-1573(84)90160-1}}.

\bibitem{Balazs84}
N.~Balazs, B.~Jennings,
  \href{http://www.sciencedirect.com/science/article/pii/0370157384901510}{Wigner's
  function and other distribution functions in mock phase spaces}, Phys. Rep.
  104 (1984) 347--391.
\newblock \href {https://doi.org/https://doi.org/10.1016/0370-1573(84)90151-0}
  {\path{doi:https://doi.org/10.1016/0370-1573(84)90151-0}}.

\bibitem{Klauder68}
J.~R. Klauder, E.~C.~G. Sudarshan, Fundamentals of Quantum Optics, Dover, New
  York, 2006.

\bibitem{GardinerQNoise}
C.~W. Gardiner, P.~Zoller, Quantum Noise: A Handbook of Markovian and
  Non-Markovian Quantum Stochastic, 3rd Edition, Springer, Berlin, 2004.

\bibitem{Englert03}
B.-G. Englert, K.~W\'{o}dkiewicz,
  \href{https://doi.org/10.1142/S0219749903000206}{Tutorial notes on one-party
  and two-party gaussian states}, Int. J. Quantum Inf. 01 (2003) 153--188.
\newblock \href {https://doi.org/10.1142/S0219749903000206}
  {\path{doi:10.1142/S0219749903000206}}.

\bibitem{Carruthers65}
P.~Carruthers, M.~M. Nieto, \href{https://doi.org/10.1119/1.1971895}{Coherent
  states and the forced quantum oscillator}, Am. J. Phys. 33 (1965) 537--544.
\newblock \href {https://doi.org/10.1119/1.1971895}
  {\path{doi:10.1119/1.1971895}}.

\bibitem{Barlow15}
T.~M. Barlow, R.~Bennett, A.~Beige,
  \href{https://doi.org/10.1080/09500340.2014.992992}{A master equation for a
  two-sided optical cavity}, J. Mod. Opt. 62 (2015) S11--S20.
\newblock \href {https://doi.org/10.1080/09500340.2014.992992}
  {\path{doi:10.1080/09500340.2014.992992}}.

\bibitem{Wilson-Rae08}
I.~Wilson-Rae, N.~Nooshi, J.~Dobrindt, T.~J. Kippenberg, W.~Zwerger,
  \href{https://doi.org/10.1088/1367-2630/10/9/095007}{Cavity-assisted
  backaction cooling of mechanical resonators}, New J. Phys. 10 (2008) 095007.
\newblock \href {https://doi.org/10.1088/1367-2630/10/9/095007}
  {\path{doi:10.1088/1367-2630/10/9/095007}}.

\bibitem{Portugal23}
P.~Portugal, F.~Brange, K.~S.~U. Kansanen, P.~Samuelsson, C.~Flindt,
  \href{https://link.aps.org/doi/10.1103/PhysRevResearch.5.033091}{Photon
  emission statistics of a driven microwave cavity}, Phys. Rev. Res. 5 (2023)
  033091.
\newblock \href {https://doi.org/10.1103/PhysRevResearch.5.033091}
  {\path{doi:10.1103/PhysRevResearch.5.033091}}.

\bibitem{McCormick19}
K.~C. McCormick, J.~Keller, D.~J. Wineland, A.~C. Wilson, D.~Leibfried,
  \href{https://doi.org/10.1088/2058-9565/ab0513}{Coherently displaced
  oscillator quantum states of a single trapped atom}, Quantum Sci. Technol. 4
  (2019) 024010.
\newblock \href {https://doi.org/10.1088/2058-9565/ab0513}
  {\path{doi:10.1088/2058-9565/ab0513}}.

\bibitem{Risken}
H.~Risken, T.~Frank, The Fokker-Planck Equation: Methods of Solution and
  Applications, 2nd Edition, Springer, Berlin, 1996.

\bibitem{Kossa76}
V.~Gorini, A.~Kossakowski, E.~C.~G. Sudarshan,
  \href{https://aip.scitation.org/doi/abs/10.1063/1.522979}{Completely positive
  dynamical semigroups of $n$-level systems}, J. Math. Phys. 17 (1976)
  821--825.
\newblock \href {https://doi.org/10.1063/1.522979}
  {\path{doi:10.1063/1.522979}}.

\bibitem{Lindblad76}
G.~Lindblad, \href{https://doi.org/10.1007/BF01608499}{On the generators of
  quantum dynamical semigroups}, Commun. Math. Phys. 48 (1976) 119--130.
\newblock \href {https://doi.org/10.1007/BF01608499}
  {\path{doi:10.1007/BF01608499}}.

\bibitem{Breuer}
H.-P. Breuer, F.~Petruccione, The Theory of Open Quantum Systems, Oxford, New
  York, 2002.

\bibitem{Lu03}
H.-X. Lu, J.~Yang, Y.-D. Zhang, Z.-B. Chen,
  \href{https://link.aps.org/doi/10.1103/PhysRevA.67.024101}{Algebraic approch
  to master equations with superoperator generators of su(1,1) and su(2) lie
  algebras}, Phys. Rev. A 67 (2003) 024101.
\newblock \href {https://doi.org/10.1103/PhysRevA.67.024101}
  {\path{doi:10.1103/PhysRevA.67.024101}}.

\bibitem{Yang03}
J.~Yang, H.-X. Lu, B.~Zhao, M.-S. Zhao, Y.-D. Zhang,
  \href{https://doi.org/10.1088/0256-307X/20/6/305}{Solution to the master
  equation of a free damped harmonic oscillator with linear driving}, Chin.
  Phys. Lett. 20 (2003) 796.
\newblock \href {https://doi.org/10.1088/0256-307X/20/6/305}
  {\path{doi:10.1088/0256-307X/20/6/305}}.

\bibitem{Tay17}
B.~A.~Tay,
  \href{https://www.sciencedirect.com/science/article/pii/S0378437116307543}{Symmetry
  of bilinear master equations for a quantum oscillator}, Physica A 468 (2017)
  578--589.
\newblock \href {https://doi.org/https://doi.org/10.1016/j.physa.2016.10.067}
  {\path{doi:https://doi.org/10.1016/j.physa.2016.10.067}}.

\bibitem{Tay17b}
B.~A.~Tay,
  \href{http://www.sciencedirect.com/science/article/pii/S0378437117301449}{Solutions
  of generic bilinear master equations for a quantum oscillator—positive and
  factorized conditions on stationary states}, Physica A 477 (2017) 42--64.
\newblock \href {https://doi.org/http://dx.doi.org/10.1016/j.physa.2017.02.020}
  {\path{doi:http://dx.doi.org/10.1016/j.physa.2017.02.020}}.

\bibitem{Tay20}
B.~A.~Tay,
  \href{http://www.sciencedirect.com/science/article/pii/S0378437120303885}{Eigenvalues
  of the liouvillian of quadratic master equation for a harmonic oscillator},
  Physica A 556 (2020) 124768.
\newblock \href {https://doi.org/https://doi.org/10.1016/j.physa.2020.124768}
  {\path{doi:https://doi.org/10.1016/j.physa.2020.124768}}.

\bibitem{Tay23}
B.~A. Tay,
  \href{https://www.sciencedirect.com/science/article/pii/S0378437123002911}{Liouvillian
  exceptional points in continuous variable system}, Physica A 620 (2023)
  128736.
\newblock \href {https://doi.org/https://doi.org/10.1016/j.physa.2023.128736}
  {\path{doi:https://doi.org/10.1016/j.physa.2023.128736}}.

\bibitem{McDonald23}
A.~McDonald, A.~A. Clerk,
  \href{https://link.aps.org/doi/10.1103/PhysRevResearch.5.033107}{Third
  quantization of open quantum systems: Dissipative symmetries and connections
  to phase-space and keldysh field-theory formulations}, Phys. Rev. Res. 5
  (2023) 033107.
\newblock \href {https://doi.org/10.1103/PhysRevResearch.5.033107}
  {\path{doi:10.1103/PhysRevResearch.5.033107}}.

\bibitem{Berry04}
M.~Berry, \href{https://doi.org/10.1023/B:CJOP.0000044002.05657.04}{Physics of
  nonhermitian degeneracies}, Czech. J. Phys, 54~(10) (2004) 1039--1047.
\newblock \href {https://doi.org/10.1023/B:CJOP.0000044002.05657.04}
  {\path{doi:10.1023/B:CJOP.0000044002.05657.04}}.

\bibitem{Heiss12}
W.~D. Heiss, \href{https://dx.doi.org/10.1088/1751-8113/45/44/444016}{The
  physics of exceptional points}, J. Phys. A: Math. Theor. 45 (2012) 444016.
\newblock \href {https://doi.org/10.1088/1751-8113/45/44/444016}
  {\path{doi:10.1088/1751-8113/45/44/444016}}.

\bibitem{Ashida21}
Y.~Ashida, Z.~Gong, M.~Ueda,
  \href{https://doi.org/10.1080/00018732.2021.1876991}{Non-hermitian physics},
  Advances in Physics 69~(3) (2020) 249--435.
\newblock \href
  {http://arxiv.org/abs/https://doi.org/10.1080/00018732.2021.1876991}
  {\path{arXiv:https://doi.org/10.1080/00018732.2021.1876991}}, \href
  {https://doi.org/10.1080/00018732.2021.1876991}
  {\path{doi:10.1080/00018732.2021.1876991}}.

\bibitem{Nori19}
F.~Minganti, A.~Miranowicz, R.~W. Chhajlany, F.~Nori,
  \href{https://link.aps.org/doi/10.1103/PhysRevA.100.062131}{Quantum
  exceptional points of non-hermitian hamiltonians and liouvillians: The
  effects of quantum jumps}, Phys. Rev. A 100 (2019) 062131.
\newblock \href {https://doi.org/10.1103/PhysRevA.100.062131}
  {\path{doi:10.1103/PhysRevA.100.062131}}.

\bibitem{Bergholtz21}
E.~J. Bergholtz, J.~C. Budich, F.~K. Kunst,
  \href{https://link.aps.org/doi/10.1103/RevModPhys.93.015005}{Exceptional
  topology of non-hermitian systems}, Rev. Mod. Phys. 93 (2021) 015005.
\newblock \href {https://doi.org/10.1103/RevModPhys.93.015005}
  {\path{doi:10.1103/RevModPhys.93.015005}}.

\bibitem{Gaidash25}
A.~Gaidash, A.~D. Kiselev, A.~Kozubov, G.~Miroshnichenko,
  \href{https://link.aps.org/doi/10.1103/2lgr-34qp}{Lindblad dynamics of open
  multimode bosonic systems: Algebra of quadratic superoperators, exceptional
  points, and speed of evolution}, Phys. Rev. A 111 (2025) 062211.
\newblock \href {https://doi.org/10.1103/2lgr-34qp}
  {\path{doi:10.1103/2lgr-34qp}}.

\bibitem{Downing23}
C.~A. Downing, A.~Vidiella-Barranco,
  \href{https://doi.org/10.1038/s41598-023-37964-7}{Parametrically driving a
  quantum oscillator into exceptionality}, Sci. Rep. 13 (2023) 11004.
\newblock \href {https://doi.org/10.1038/s41598-023-37964-7}
  {\path{doi:10.1038/s41598-023-37964-7}}.

\bibitem{Mylnikov25}
V.~Y. Mylnikov, S.~O. Potashin, G.~S. Sokolovskii, N.~S. Averkiev,
  \href{https://link.aps.org/doi/10.1103/PhysRevResearch.7.013061}{Emergent
  equilibrium and quantum criticality in a two-photon dissipative oscillator},
  Phys. Rev. Res. 7 (2025) 013061.
\newblock \href {https://doi.org/10.1103/PhysRevResearch.7.013061}
  {\path{doi:10.1103/PhysRevResearch.7.013061}}.

\bibitem{Weedbrook12}
C.~Weedbrook, S.~Pirandola, R.~Garc\'{\i}a-Patr\'on, N.~J. Cerf, T.~C. Ralph,
  J.~H. Shapiro, S.~Lloyd,
  \href{https://link.aps.org/doi/10.1103/RevModPhys.84.621}{Gaussian quantum
  information}, Rev. Mod. Phys. 84 (2012) 621--669.
\newblock \href {https://doi.org/10.1103/RevModPhys.84.621}
  {\path{doi:10.1103/RevModPhys.84.621}}.

\bibitem{Walter14}
S.~Walter, A.~Nunnenkamp, C.~Bruder,
  \href{https://link.aps.org/doi/10.1103/PhysRevLett.112.094102}{Quantum
  synchronization of a driven self-sustained oscillator}, Phys. Rev. Lett. 112
  (2014) 094102.
\newblock \href {https://doi.org/10.1103/PhysRevLett.112.094102}
  {\path{doi:10.1103/PhysRevLett.112.094102}}.

\bibitem{Li25}
Y.~Li, Z.~Xie, X.~Yang, Y.~Li, X.~Zhao, X.~Cheng, X.~Peng, J.~Li, E.~Lutz,
  Y.~Lin, J.~Du,
  \href{https://www.science.org/doi/abs/10.1126/sciadv.ady5649}{Experimental
  realization and synchronization of a quantum van der pol oscillator}, Sci.
  Adv. 11~(41) (2025) eady5649.
\newblock \href {https://doi.org/10.1126/sciadv.ady5649}
  {\path{doi:10.1126/sciadv.ady5649}}.

\bibitem{Tay04}
B.~A. Tay,
  \href{https://repositories.lib.utexas.edu/items/21c16548-d10b-4a37-99ae-25b6cb326152}{Coherence
  and Decoherence Processes of a Harmonic Oscillator Coupled with Finite
  Temperature Field -- Exact Eigenbasis Solution of Kossakowski-Lindblad's
  Equation, Ph.D. Dissertation}, University of Texas at Austin, 2004.

\bibitem{Tay07}
B.~A. Tay, T.~Petrosky,
  \href{http://link.aps.org/doi/10.1103/PhysRevA.76.042102}{Thermal symmetry of
  the markovian master equation}, Phys. Rev. A 76 (2007) 042102.
\newblock \href {https://doi.org/10.1103/PhysRevA.76.042102}
  {\path{doi:10.1103/PhysRevA.76.042102}}.

\bibitem{Vazquez18}
P.~C. L\'opez~V\'azquez,
  \href{https://link.aps.org/doi/10.1103/PhysRevA.98.042128}{Driving a
  dissipative quantum oscillator}, Phys. Rev. A 98 (2018) 042128.
\newblock \href {https://doi.org/10.1103/PhysRevA.98.042128}
  {\path{doi:10.1103/PhysRevA.98.042128}}.

\bibitem{Ma18}
S.~Ma, M.~J. Woolley, I.~R. Petersen, A derivation of moment evolution
  equations for linear open quantum systems, in: 2018 33rd Youth Academic
  Annual Conference of Chinese Association of Automation (YAC), 2018, pp.
  6--11.
\newblock \href {https://doi.org/10.1109/YAC.2018.8405797}
  {\path{doi:10.1109/YAC.2018.8405797}}.

\bibitem{Tay08}
B.~A. Tay, T.~Petrosky,
  \href{http://dx.doi.org/doi/10.1063/1.3005968}{Biorthonormal eigenbasis of a
  markovian master equation for the quantum brownian motion}, J. Math. Phys. 49
  (2008) 113301.
\newblock \href {https://doi.org/DOI:10.1063/1.3005968}
  {\path{doi:DOI:10.1063/1.3005968}}.

\bibitem{Tay19b}
B.~A. Tay,
  \href{http://www.sciencedirect.com/science/article/pii/S0378437119306752}{Damping
  modes of harmonic oscillator in open quantum systems}, Physica A 526 (2019)
  121119.
\newblock \href {https://doi.org/https://doi.org/10.1016/j.physa.2019.121119}
  {\path{doi:https://doi.org/10.1016/j.physa.2019.121119}}.

\bibitem{CL83}
A.~Caldeira, A.~Leggett,
  \href{http://www.sciencedirect.com/science/article/pii/0378437183900134}{Path
  integral approach to quantum brownian motion}, Physica A 121 (1983) 587--616.
\newblock \href {https://doi.org/https://doi.org/10.1016/0378-4371(83)90013-4}
  {\path{doi:https://doi.org/10.1016/0378-4371(83)90013-4}}.

\bibitem{Passante95}
R.~Passante, T.~Petrosky, I.~Prigogine,
  \href{https://www.sciencedirect.com/science/article/pii/037843719500147Y}{Long-time
  behaviour of self-dressing and indirect spectroscopy}, Physica A 218 (1995)
  437--456.
\newblock \href {https://doi.org/https://doi.org/10.1016/0378-4371(95)00147-Y}
  {\path{doi:https://doi.org/10.1016/0378-4371(95)00147-Y}}.

\bibitem{HPZ92}
B.~L. Hu, J.~P. Paz, Y.~Zhang,
  \href{http://link.aps.org/doi/10.1103/PhysRevD.45.2843}{Quantum brownian
  motion in a general environment: Exact master equation with nonlocal
  dissipation and colored noise}, Phys. Rev. D 45 (1992) 2843--2861.
\newblock \href {https://doi.org/10.1103/PhysRevD.45.2843}
  {\path{doi:10.1103/PhysRevD.45.2843}}.

\bibitem{Barsegov02}
T.~Petrosky, V.~Barsegov,
  \href{http://link.aps.org/doi/10.1103/PhysRevE.65.046102}{Quantum
  decoherence, zeno process, and time symmetry breaking}, Phys. Rev. E 65
  (2002) 046102.
\newblock \href {https://doi.org/10.1103/PhysRevE.65.046102}
  {\path{doi:10.1103/PhysRevE.65.046102}}.

\bibitem{Wigner32}
E.~Wigner, \href{https://link.aps.org/doi/10.1103/PhysRev.40.749}{On the
  quantum correction for thermodynamic equilibrium}, Phys. Rev. 40 (1932)
  749--759.
\newblock \href {https://doi.org/10.1103/PhysRev.40.749}
  {\path{doi:10.1103/PhysRev.40.749}}.

\bibitem{Prigogine73}
I.~Prigogine, C.~George, F.~Henin, L.~Rosenfeld, A unified formulation of
  dynamics and thermodynamics, Chem. Scr. 4 (1973) 5--32.

\bibitem{Talkner81}
P.~Talkner, \href{http://dx.doi.org/10.1007/BF01307328}{Gauss markov process of
  a quantum oscillator}, Z. Phy. B: Cond. Matt. 41~(4) (1981) 365--374.
\newblock \href {https://doi.org/10.1007/BF01307328}
  {\path{doi:10.1007/BF01307328}}.

\bibitem{Gilmore}
R.~Gilmore, Lie Groups, Lie Algebras, and Some of Their Applications, John
  Wiley \& Sons, New York, 1974.

\bibitem{Simon88}
R.~Simon, E.~C.~G. Sudarshan, N.~Mukunda,
  \href{https://link.aps.org/doi/10.1103/PhysRevA.37.3028}{Gaussian pure states
  in quantum mechanics and the symplectic group}, Phys. Rev. A 37 (1988)
  3028--3038.
\newblock \href {https://doi.org/10.1103/PhysRevA.37.3028}
  {\path{doi:10.1103/PhysRevA.37.3028}}.

\bibitem{Tay19a}
B.~A. Tay, \href{https://doi.org/10.1140/epjb/e2019-90630-0}{General symmetry
  in the reduced dynamics of two-level system}, Eur. Phys. J. B 92~(4) (2019)
  84.
\newblock \href {https://doi.org/10.1140/epjb/e201i9-90630-0}
  {\path{doi:10.1140/epjb/e201i9-90630-0}}.

\bibitem{Magnus54}
W.~Magnus,
  \href{https://onlinelibrary.wiley.com/doi/abs/10.1002/cpa.3160070404}{On the
  exponential solution of differential equations for a linear operator},
  Commun. Pure and App. Math. 7 (1954) 649--673.
\newblock \href {https://doi.org/https://doi.org/10.1002/cpa.3160070404}
  {\path{doi:https://doi.org/10.1002/cpa.3160070404}}.

\bibitem{Briegel93}
H.-J. Briegel, B.-G. Englert,
  \href{https://link.aps.org/doi/10.1103/PhysRevA.47.3311}{Quantum optical
  master equations: The use of damping bases}, Phys. Rev. A 47 (1993)
  3311--3329.
\newblock \href {https://doi.org/10.1103/PhysRevA.47.3311}
  {\path{doi:10.1103/PhysRevA.47.3311}}.

\bibitem{Honda10}
D.~Honda, H.~Nakazato, M.~Yoshida,
  \href{http://dx.doi.org/doi/10.1063/1.3442363}{Spectral resolution of the
  liouvillian of the lindblad master equation for a harmonic oscillator},
  J.~Math.~Phys. 51~(7) (2010) 072107.
\newblock \href {https://doi.org/DOI:10.1063/1.3442363}
  {\path{doi:DOI:10.1063/1.3442363}}.

\bibitem{Heiss16}
D.~Heiss, \href{https://doi.org/10.1038/nphys3864}{Circling exceptional
  points}, Nat. Phys. 12 (2016) 823--824.
\newblock \href {https://doi.org/10.1038/nphys3864}
  {\path{doi:10.1038/nphys3864}}.

\bibitem{Arfken}
G.~B. Arfken, H.~J. Weber, F.~E. Harris, Mathematical Methods for Physicists,
  7th Edition, Academic Press, Boston, 2012.

\end{thebibliography}
\end{document}